\PassOptionsToPackage{table,xcdraw,dvipsnames,x11names}{xcolor}
\documentclass[sigconf, nonacm]{acmart}

\makeatletter
\apptocmd{\@mkauthors@iii}{%
  \global\setbox\mktitle@bx=\vbox{%
    \unvbox\mktitle@bx
    \vskip 10pt
  }%
}{}{}
\makeatother
\renewcommand\footnotetextcopyrightpermission[1]{}
\usepackage{amsmath,amsfonts}
\usepackage{circledsteps}

\usepackage[linesnumbered,ruled,vlined]{algorithm2e}
\SetInd{0.3em}{0.8em}
\SetVlineSkip{1pt}

\SetAlFnt{\small}            
\SetAlCapFnt{\small}         
\SetAlCapNameFnt{\small}     
\SetCommentSty{scriptsize}   

\SetKwInput{KwInput}{Input}
\SetKwInput{KwOutput}{Output}
\SetKwInput{KwVar}{Variables}
\SetKw{Return}{return}
\SetEndCharOfAlgoLine{}

\SetKwComment{com}{\hfill $\triangleright$\scriptsize\space}{} 
\SetKwComment{sep}{\hfill }{}

\usepackage{stfloats}
\usepackage{xcolor}
\usepackage{graphicx}
\usepackage{textcomp}
\usepackage{mwe}
\usepackage{bm}

\usepackage[dvipsnames, x11names]{xcolor}
\def\BibTeX{{\rm B\kern-.05em{\sc i\kern-.025em b}\kern-.08em
    T\kern-.1667em\lower.7ex\hbox{E}\kern-.125emX}}

\usepackage[most]{tcolorbox}
\usepackage{xparse}
\usepackage{lipsum}
\usepackage{pdfpages}
\usepackage{footnote}
\usepackage{refcount}
\usepackage{enumitem}
\usepackage{hyperref}
\usepackage{float}
\usepackage{multirow}
\usepackage{algpseudocode}
\usepackage{wrapfig}
\usepackage{todonotes}

\usepackage{booktabs}
\usepackage{array}

\definecolor{brown-web}{rgb}{0.65, 0.16, 0.16}
\definecolor{oliv-green}{RGB}{60, 128, 49}
\definecolor{light-pink}{RGB}{251, 229, 229}
\definecolor{light-orange}{RGB}{255, 213, 128}
\newcommand{\blue}[1]{\textcolor{blue}{#1}}
\newcommand{\shaon}[1]{\textcolor{red}{#1}}
\newcommand{\red}[1]{\textcolor{red}{#1}}
\newcommand{\ex}[1]{\textcolor{brown-web}{#1}}
\newcommand{\sh}[1]{\textcolor{oliv-green}{#1}}

\usepackage{soul}
\sethlcolor{yellow} 

\soulregister\cite7 
\soulregister\ref7 
\soulregister\pageref7 
\soulregister{\shaon}{1}
\soulregister{\red}{1}
\soulregister{\ex}{1}
\soulregister{\sh}{1}
\soulregister{\emph}{1}

\makeatletter
\newcommand{\stitle}[1]{\noindent\textbf{#1\@addpunct{.}}}
\newcommand{\ititle}[1]{\noindent\emph{#1\@addpunct{.}}}
\makeatother

\begin{document}
\title{Discovery-Driven Integration of Disjoint Tables via Text}

\author{Md Ataur Rahman}
\affiliation{%
  \institution{UPC, BarcelonaTech}
  \city{Barcelona}
  \country{Spain}
}
\email{md.ataur.rahman@upc.edu}

\author{Dimitris Sacharidis}
\affiliation{%
  \institution{\mbox{Université Libre de Bruxelles}}
  \city{Brussels}
  \country{Belgium}
}
\email{dimitris.sacharidis@ulb.be}

\author{Oscar Romero}
\affiliation{%
  \institution{UPC, BarcelonaTech}
  \city{Barcelona}
  \country{Spain}
}
\email{oscar.romero@upc.edu}

\author{Sergi Nadal}
\affiliation{%
  \institution{UPC, BarcelonaTech}
  \city{Barcelona}
  \country{Spain}
}
\email{sergi.nadal@upc.edu}


\begin{abstract}
Integrating heterogeneous datasets within data lakes is a critical challenge, particularly for semantically related tables that lack the explicit attributes needed to be joined. We study \textbf{Discovery-Driven Integration}, where the relevant sources and their missing relational structure must be discovered before integration. In this setting, unstructured text provides the evidence that connects otherwise disjoint tables. The fundamental challenge is to discover the relationships at a fine-grained level that connect individual rows from different tables through specific sentences. We formalize this task as \textbf{Text-Mediated Join Path Discovery} and propose a horizontal bidirectional cross-attention architecture called \textsf{\textbf{LOKI}} (\textsf{\textbf{L}}atent-space \textsf{\textbf{O}}ptimization for \textsf{\textbf{K}}nowledge \textsf{\textbf{I}}ntegration)\footnote{\texttt{\textbf{\ex{Code \& Artifacts}}}: \blue{\url{https://github.com/dtim-upc/LOKI}}} that learns contextualized representations of table rows and sentences. Through a global table-text contrastive objective, fine-grained row-sentence associations emerge without explicit local supervision. Existing multi-modal discovery methods largely retrieve coarse-grained column-text associations, whereas integration systems assume supplied row-text links, schemas, or queries. LOKI instead transforms these implicit associations into explicit, interpretable join paths, organizes them into relation-consistent groups, and materializes them as typed integrated tables with sentence-level provenance. Comprehensive evaluations on real-world benchmarks demonstrate that LOKI consistently outperforms state-of-the-art multi-modal data discovery approaches, and materializes typed integrated tables with 0.982 macro typed-pair precision while being up to 40 times cheaper in LLM API cost than direct prompting.
\end{abstract}

\maketitle


\section{Introduction}

Data lakes serve as centralized repositories consolidating large volumes of heterogeneous data, often containing thousands of datasets with structured tables and large collections of unstructured text documents. Information embedded in text contains valuable insights that remain inaccessible unless extracted and integrated with structured data: in a hospital data lake, for example, discharge summaries may link patient records to medication treatments. Extracting and integrating such information makes it available for analysis.

Recently, several methods have been proposed for \textbf{text-table querying}, allowing users to retrieve information from both modalities, e.g., \cite{UrbanB24, JoT24, RahmanN0S24, 00020F25}. However, these methods share a key limitation: they assume that the \emph{integration} between the two modalities is already known. In particular, they require knowledge of (1) the \emph{row-text links}, i.e., which part of the text is associated to each table row, and (2) the \emph{text schema}, i.e., what information is contained in this part of the text. Essentially, they perform query-time extraction over a predefined text-table integration. For example, ELEET~\cite{UrbanB24} processes queries that specify a \emph{latent attribute} to be extracted from text for each table row, while ThalamusDB~\cite{JoT24} processes queries containing a Boolean \emph{predicate} expressed in natural language over a text document. The latent attribute in the former and the predicate in the latter specify text schema, while the association between text and table rows is assumed to be known.

\medskip

\begin{figure*}[t]
\centering
\includegraphics[width=1\linewidth]{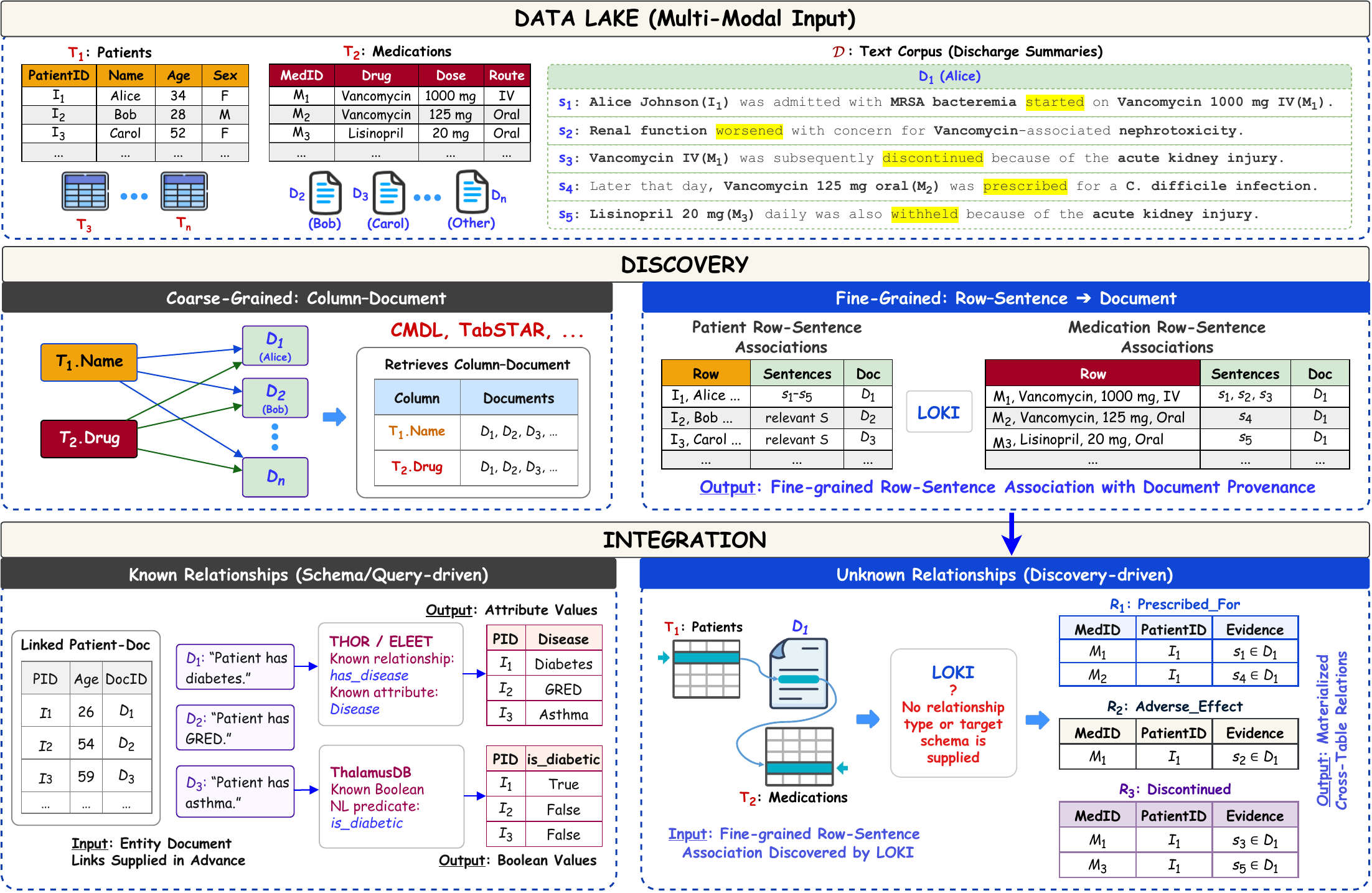}
\vspace{-4mm}
\caption{Discovery-driven integration within a multi-modal data lake. Coarse-grained approaches retrieve table-text pairs, and integration systems require supplied row-text links and predefined text schema. LOKI instead derives document relevance from fine-grained row-sentence associations and uses them to disambiguate and materialize evidence-backed cross-table relations without a predefined relationship type or target schema.}
\label{fig:motivating_example_loki}
\end{figure*}

\begin{tcolorbox}[boxrule=0.2pt, boxsep=0pt, colback=Snow1, before skip=\smallskipamount, after skip=\smallskipamount]
\stitle{Motivating Example} Consider the hospital data lake in Fig.~\ref{fig:motivating_example_loki}, with tables such as the \ex{\textit{Patients}} table \ex{\textit{$T_1$}} and the \ex{\textit{Medications}} table \ex{\textit{$T_2$}}, and text documents such as the \ex{\textit{Discharge Summaries}} \ex{\textit{$D_1$}} and \ex{\textit{$D_2$}}. Assume that (1) \ex{\textit{$D_1$}} concerns patient \ex{\textit{$I_1$}} and \ex{\textit{$D_2$}} concerns patient \ex{\textit{$I_2$}}, and (2) \ex{\textit{Diagnosis}} information appears in the discharge summaries and must be extracted per patient. Under these assumptions, the \ex{\textit{Diagnosis}} table \ex{\textit{$T_a$}} can be extracted from the text and linked to the \ex{\textit{Patients}} table \ex{\textit{$T_1$}}, enabling querying across the text and tables.
\end{tcolorbox}

\medskip

%

To fully extract value from data lakes, we must remove these two assumptions and enable \textbf{text-table integration}, where neither row-text associations nor text schemas are known beforehand. Existing multi-modal discovery and retrieval techniques \cite{FernandezMQEIMO18, FernandezM19, AnadiotisBCGHMM22, EltabakhKEA23} identify only \emph{coarse-grained} associations between columns (or tables) and text, missing \emph{fine-grained} row-text links. This is insufficient for integration in two ways. It can collapse semantically distinct relationships: a discharge summary may state that a patient was prescribed aspirin, had an adverse reaction, and later discontinued it (three relationships a coarse-grained association cannot distinguish). It can also introduce spurious ones (e.g., linking a patient to every medication merely discussed rather than prescribed).

\medskip

\begin{tcolorbox}[boxrule=0.2pt, boxsep=0pt, colback=Snow1, before skip=\smallskipamount, after skip=\smallskipamount]
\stitle{Running Example} On the example above, coarse-grained discovery may identify, for instance, that the \ex{\textit{$T_1.Name$}} column of the \ex{\textit{Patients}} table \ex{\textit{$T_1$}} is related to the discharge summaries \ex{\textit{$D_1$}} and \ex{\textit{$D_2$}}. Fine-grained discovery, in contrast, identifies row-text links, such as the links between the first row of \ex{\textit{$T_1$}}, corresponding to patient \ex{\textit{$I_1$}}, and sentences \ex{\textit{$s_1$-- $s_5$}} in \ex{\textit{$D_1$}}.
\end{tcolorbox}

\medskip

To the best of our knowledge, no existing method automatically discovers both the fine-grained row-text associations and the schemas needed to integrate text with existing tables. In this work, we present LOKI, a system that discovers such associations, starting from fine-grained \emph{row-text links}: two links with overlapping text spans and rows from different tables indicate that the text expresses a relationship between them. We call the resulting sequence of a table row, a supporting text span, and another table row a \emph{text-mediated join path}. LOKI collects the join paths for each pair of tables, examining their supporting text to determine whether they express one relationship or several, each defining a \emph{text schema}.


\vspace{4mm}
\stitle{\ex{Overview of LOKI}}
At the heart of LOKI is a cross-modal encoder that takes a table and a text document as input and produces embeddings for each table row and document sentence. The encoder combines a frozen text encoder with a trainable bidirectional attention mechanism that contextualizes each row embedding with information from the document and each sentence embedding with information from the table: rows attend to sentences in the forward direction, and sentences attend to rows in the reverse. LOKI uses the resulting contextualized embeddings to compute a row-sentence pair-score matrix that captures fine-grained associations between rows and sentences. LOKI does not require costly row-sentence annotations for training. Instead, it relies on a small number of potentially noisy coarse table-document annotations, which can be obtained from weak signals such as syntactic or embedding-based similarity, or weak-supervision techniques~\cite{RatnerSWSR16}, as done in~\cite{YinNYR20,EltabakhKEA23}.

LOKI trains the encoder contrastively on triples $(T, D^+, D^-)$ with two objectives: a global one that ranks $D^+$ above $D^-$ for table $T$ from the aggregated pair scores, and a local one that, drawing on multiple-instance learning~\cite{IlseTW18,MiechASLSZ20}, separates a sampled row's best-matching sentence in $D^+$ from its best in $D^-$. Together they induce fine-grained associations from coarse supervision. Once trained, LOKI discovers row-text links and text-mediated join paths: for each document, it identifies coarsely related tables, considers pairs of them, and uses the overlapping supporting text spans of their row-text links to construct join paths. LOKI encodes and groups the resulting paths using density-based clustering, where each cluster represents a distinct relationship between a pair of tables. An LLM names each relationship based on the evidence sentences, and LOKI creates one table per cluster, populated with the join paths.

\begin{tcolorbox}[boxrule=0.2pt, boxsep=0pt, colback=Snow1, before skip=\smallskipamount, after skip=\smallskipamount]
\stitle{Running Example} Continuing the example, fine-grained discovery identifies that sentences such as \ex{\textit{$s_1$}} in \ex{\textit{$D_1$}} relate patient \ex{\textit{$I_1$}} in \ex{\textit{$T_1$}} and medication \ex{\textit{$M_1$}} in \ex{\textit{$T_2$}}, revealing the text-mediated join path \ex{\textit{$(I_1, D_1.s_1, M_1)$}}: a relationship between \ex{\textit{$I_1$}} and \ex{\textit{$M_1$}} mediated by \ex{\textit{$D_1.s_1$}}. LOKI collects such join paths per table pair and groups them by the relationships expressed in their supporting text. For \ex{\textit{$T_1$}} and \ex{\textit{$T_2$}}, this reveals three relationships between patients and medications: \ex{\textit{$(I_1,D_1.s_1,M_1)$}} expresses a \ex{\textit{Prescribed\_For}}, \ex{\textit{$(I_1,D_1.s_2,M_1)$}} an \ex{\textit{Adverse\_Effect}}, and \ex{\textit{$(I_1,D_1.s_3,M_1)$}} a \ex{\textit{Discontinued}} relationship. LOKI therefore creates three distinct text schemas, one per relationship, and populates the tables with the discovered instances.
\end{tcolorbox}

\medskip

\stitle{\ex{Contributions}} This work makes the following contributions to multi-modal discovery and querying:

\begin{itemize}[leftmargin=3mm]
\setlength\itemsep{1mm}

\item \textbf{Novel problem.} We introduce Discovery-Driven Integration, the setting in which relationships between tables mediated by text documents are unknown, and formalize its core task as Text-Mediated Join Path Discovery. To the best of our knowledge, no prior method jointly discovers, disambiguates, and materializes latent row-sentence-row relationships, into integrated tables.

\item \textbf{Joint representation-learning architecture.} We introduce the LOKI encoder, which learns contextualized row-text representations from coarse table-text supervision alone, enabling fine-grained associations to emerge without explicit row-sentence annotations. Unlike existing column-based approaches that use vertical attention within a table column~\cite{YinNYR20,EltabakhKEA23}, LOKI uses horizontal bidirectional cross-attention between rows and sentences.

\item \textbf{SOTA Performance.} We evaluate LOKI on an end-to-end integration task using real-world datasets, bridging disjoint tables through fine-grained row-sentence-row evidence and materializing typed relations at a fraction of the LLM token cost. LOKI also outperforms existing cross-modal discovery and representation-learning methods at coarse-grained discovery.
\end{itemize}
\section{Related Work}

We situate LOKI at the intersection of multi-modal data discovery and integration: existing discovery systems retrieve related tables or table-document pairs, while integration systems operate over predefined links, schemas, or queries. LOKI bridges the two through \textit{discovery-driven integration}, using text to discover and materialize relationships between disjoint tables.

\medskip

\stitle{Join-Path Discovery}
Our definition of \textit{Text-Mediated Join Path} introduces a row-to-row association mediated by natural language sentences, whereas prior work defined join paths primarily at the table or schema level. In systems such as \textbf{Metam}~\cite{GalhotraGF23}, \textbf{Data Civilizer}~\cite{MansourDFQTAEIM18}, \textbf{Aurum}~\cite{FernandezAKYMS18}, and \textbf{SemDisc}~\cite{mohammad2026qualitative}, join paths denote ordered chains of joinable datasets used for data discovery or augmentation, but these remain purely structural, without textual mediation or record-level grounding. \textbf{Ver}~\cite{GongZGF23,DharmawanKGYZGK24} extends this notion to schema-agnostic view discovery with no predefined foreign keys. The novelty of our formulated \textit{join path} lies in its formulation as an explicit fine-grained \textit{transitive bridge} through mediating sentences as evidence of latent relationships. This also differs from multi-modal discovery systems that stop at one-hop table-document retrieval~\cite{EltabakhKEA23}.

\medskip

\stitle{Semantic Table Discovery}
Current research moves beyond syntactic value overlaps to capture semantic relatedness between tables. \textbf{DeepJoin}~\cite{Dong0NEO23} and \textbf{Snoopy}~\cite{GuoMHCG25} fine-tune Transformer-based models to encode column values, identifying joinable columns based on semantic proximity rather than exact matches. \textbf{FREYJA}~\cite{MaynouFREYJA24} combines data profiling with predictive modeling to approximate semantics-aware joins, while \textbf{OmniMatch}~\cite{Koutras2025} constructs a graph of columns to propagate similarity signals. \textbf{SANTOS}~\cite{KhatiwadaFSCGMR23} retrieves unionable tables using semantic relationships between column pairs from an external or synthesized knowledge base, while \textbf{KGLiDS}~\cite{HelaliMVCHCAH024} captures dataset semantics with learned data profiles and knowledge graphs. Despite their different mechanisms, these approaches derive their discovery signals from table values, profiles, metadata, or semantics constructed from tabular artifacts, and cannot discover relationships between disjoint tables when the connecting evidence exists only in external text, which \textbf{LOKI} instead exploits as a semantic bridge to discover row-level relationships.

\medskip

\stitle{Joint Representation Learning}
To bridge the gap between tables and text, prior work learns a unified embedding space. \textbf{SEMPROP} \cite{FernandezMQEIMO18} and \textbf{Termite} \cite{FernandezM19} project structured and unstructured data into a shared latent space to retrieve related items based on vector proximity. \textbf{TABERT} \cite{YinNYR20} and \textbf{PNEUMA} \cite{BalakaAWGKF25} linearize tables to learn joint representation with text, often for Question Answering (QA) or Retrieval-Augmented Generation (RAG) tasks. These approaches operate at a coarse schema or document granularity: they function as semantic search engines rather than integration systems and do not extract explicit, row-level join paths, whereas \textbf{LOKI} resolves fine-grained row-sentence associations, transforming black-box retrieval scores into interpretable, transitive join paths.

\medskip

\stitle{Multi-Modal Data Discovery}
Several systems retrieve related data across structured and unstructured data. \textbf{OTTER}~\cite{HuangZLGJD22} jointly retrieves tables and text, but its retrieval unit is a table row pre-linked to text passages via entity linking, relying on existing links rather than discovering them. \textbf{CMDL}~\cite{EltabakhKEA23} and \textbf{Tri-Encoder}~\cite{kostic21} support data-lake queries over structured and unstructured data, returning ranked table- or column-document associations. \textbf{TabSTAR}~\cite{tabSstar2025} jointly represents structured attributes and text fields for tabular prediction, also supporting cross-modal retrieval. These systems narrow the candidate sources, but their coarse-grained outputs do not determine which rows connect, which sentences support the connection, or which relationships they express; using them directly for integration may produce spurious joins, e.g., linking a patient to every medication in the same summary. \textbf{LOKI} instead composes fine-grained row-sentence associations into evidence-backed row-sentence-row paths for relation-specific integration.

\medskip

\stitle{Schema/Query-Driven Integration}
Existing multi-modal integration systems assume a predefined relational structure. \textbf{THOR}~\cite{RahmanN0S24} populates values for a target concept from related text. \textbf{ELEET}~\cite{UrbanB24} extracts registered attributes into latent tables and composes them with relational operators. Its multi-modal join assumes table tuples are already linked to documents. 
\textbf{ThalamusDB}~\cite{JoT24} evaluates explicit natural-language predicates within SQL queries over a known multi-modal schema, whereas \textbf{iDataLake}~\cite{00020F25} uses an analytical query to select sources and construct an execution plan. \textbf{ConnectionLens}~\cite{AnadiotisBCGHMM22} integrates heterogeneous sources by extracting and linking entities from text and other sources. These systems operate once the required links, schemas, or queries are supplied, but do not determine which cross-table relations should exist when both connections and relationship types are unknown. \textbf{LOKI}, in contrast, discovers row-sentence-row paths, disambiguates their semantics, and materializes them as evidence-backed tables without a predefined schema.


\section{The LOKI Encoder}

\begin{figure*}[t]
\centering
\includegraphics[width=0.85\linewidth]{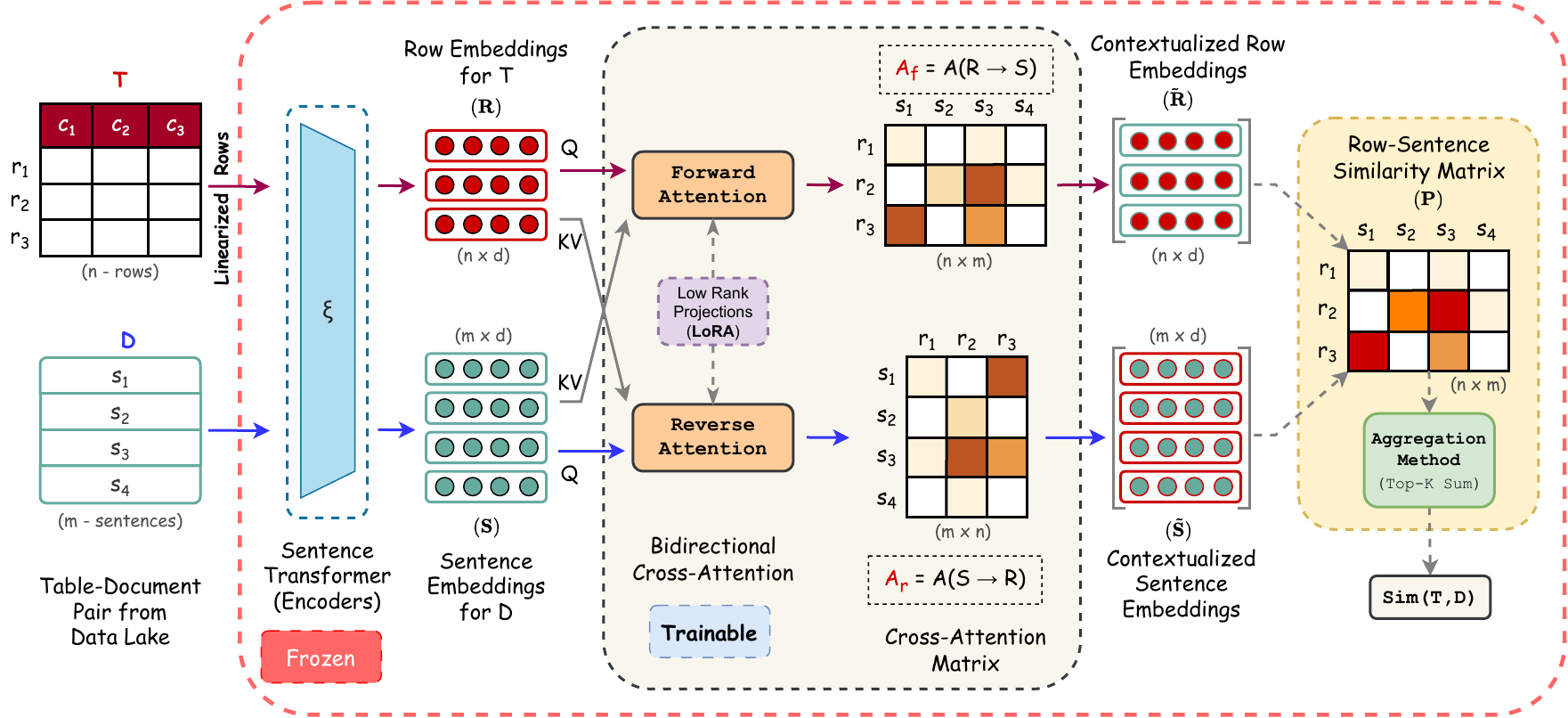}
\caption{LOKI Encoder for multi-modal representation learning. Given a candidate table-document pair $(T,D)$, the frozen encoder $\xi$ embeds the linearized table rows and document sentences as $\mathbf{R}$ and $\mathbf{S}$. Bidirectional cross-attention, parameterized by shared LoRA-adapted projections, produces contextualized embeddings $\widetilde{\mathbf{R}}$ and $\widetilde{\mathbf{S}}$ and the attention distributions $\mathbf{A}_f$ and $\mathbf{A}_r$. Their cosine similarities form the pair-score matrix $\mathbf{P}$, whose strongest entries capture fine-grained row-sentence associations and whose Top-$K$ aggregation yields the global similarity $\operatorname{Sim}(T,D)$.}
\label{fig:pipeline_loki}
\end{figure*}

We present the LOKI encoder (Fig. ~\ref{fig:pipeline_loki}) and discuss its input, its contextualization of rows and sentences, its output, and its training.

\subsection{Input and Initial Embeddings}
The LOKI encoder takes as input a text $D$ and a table $T$, where $T$ is converted into text by a row-based linearization. Specifically, each row $r_i$ from table $T$ is first transformed into a natural language string. The linearization concatenates the column names and their cell values for that row:
\begin{equation*}
\operatorname{linearize}(T) = \left[\text{``}c_1{:}\,v_{i1}; c_2{:}\,v_{i2}; \cdots ; c_{\ell}{:}\,v_{i\ell}\text{''}\right]_{i=1}^{n}
\end{equation*}
where $\ell$ is the number of columns of $T$. 
For instance, the first row of the \ex{\textit{Patients}} table \ex{$T_1$} is serialized into the string: ``\ex{\textit{PatientID: $I_1$; Name: Alice; Age: 34; Sex: F}}''. This lets LOKI treat structured data as natural language, consumable by attention modules alongside documents. LOKI feeds text $D$ and linearized table $T$ into a pre-trained sentence encoder, $\xi(\cdot)$. While token-level encoders return a contextualized vector per token, sentence encoders such as SBERT~\cite{ReimersG19} aggregate these into a single fixed-dimensional vector per linearized row and sentence:
\begin{equation*}
\resizebox{\linewidth}{!}{$
\mathbf{R} = \left[ \xi(\operatorname{linearize}(T)) \right]^{\top} \in \mathbb{R}^{n \times d}; \quad
\mathbf{S} = [\xi(s_1), \ldots, \xi(s_m)]^{\top} \in \mathbb{R}^{m \times d}
$}
\end{equation*}
Here $n$ is the number of rows of $T$, $m$ the number of sentences in $D$, and $d$ the embedding dimension of $\xi$. This yields two matrices $\mathbf{R}$ and $\mathbf{S}$ of initial, \textit{non-contextualized} embeddings that capture the general meaning of each table row (e.g., the record for \ex{\textit{Alice}}) and each sentence in isolation, without awareness of the other modality. The weights of the base encoder $\xi$ are kept \emph{frozen} to leverage its general-purpose linguistic knowledge.

\subsection{Contextualized Embeddings}
The distinguishing characteristic of the LOKI encoder is its bidirectional cross-attention module. Unlike traditional unidirectional attention, it is designed around the premise that cross-modal association requires a symmetric exchange of information: a table row's meaning is grounded by the sentences that mention its entities, and a sentence's relevance is determined by the referenced tabular data. This stage produces \emph{contextualized cross-modal representations}.

The module consists of two parallel cross-attention blocks, stabilized by a \emph{gated attention mechanism}~\cite{abs-2505-06708} that helps prevent attention collapse. We retain the Forward Attention Matrix $\mathbf{A}_f$ and Reverse Attention Matrix $\mathbf{A}_r$ as the internal attention distributions of these blocks. For a given set of queries ($\mathbf{Q}$), keys ($\mathbf{K}$), and values ($\mathbf{V}$), the attention matrix $\mathbf{A}$ and its output $\mathbf{H}$ are computed as:
\begin{gather*}
\mathbf{A}(\mathbf{Q},\mathbf{K})
=
\operatorname{softmax}\left(
\frac{
(\mathbf{Q}\mathbf{W}_Q)
(\mathbf{K}\mathbf{W}_K)^{\top}
}{
\sqrt{d_k}
}
\right);
\\
\mathbf{H}(\mathbf{Q},\mathbf{K},\mathbf{V})
=
\mathbf{A}(\mathbf{Q},\mathbf{K})
(\mathbf{V}\mathbf{W}_V)
\end{gather*}
where $\mathbf{W}_Q$, $\mathbf{W}_K$, and $\mathbf{W}_V$ are the query, key, and value projections, shared between the forward and reverse cross-attention blocks, and $d_k$ is the dimensionality of the projected keys. For parameter-efficient adaptation, we apply Low-Rank Adaptation (LoRA) to the shared projections while keeping the base encoder $\xi$ frozen, so cross-attention learns task-specific row-sentence interactions without modifying the encoder's general-purpose representations.

\medskip 

\stitle{Forward Attention (Rows $\to$ Sentences)} As depicted in Fig.~\ref{fig:pipeline_loki} (top), table rows act as queries to seek information from sentences, which serve as keys and values. This produces contextualized row representations $\widetilde{\mathbf{R}}$, enriched by the relevant sentence context:
\begin{equation*}
\mathbf{A}_f
=
\mathbf{A}(\mathbf{R},\mathbf{S});
\quad
\mathbf{H}_f
=
\mathbf{H}(\mathbf{R},\mathbf{S},\mathbf{S});
\quad
\widetilde{\mathbf{R}}
=
\mathbf{R}
+
\mathbf{H}_f
\odot
\sigma(\mathbf{R}\mathbf{W}_g^f)
\end{equation*}
This operation contextualizes each row using the sentences most relevant to it. In our running example, the \ex{\textit{Alice}} row attends most strongly to the sentence \ex{$s_1$} describing her \ex{\textit{Vancomycin}} prescription, producing a sentence-aware representation in $\widetilde{\mathbf{R}}$. The residual connection preserves the row's original identity.

\medskip

\stitle{Reverse Attention (Sentences $\to$ Rows)} As depicted in Fig.~\ref{fig:pipeline_loki} (bottom), it uses sentences as queries to attend to the table rows. This process produces the contextualized sentence representations $\widetilde{\mathbf{S}}$, enriched by the relevant row context:
\begin{equation*}
\mathbf{A}_r
=
\mathbf{A}(\mathbf{S},\mathbf{R});
\quad
\mathbf{H}_r
=
\mathbf{H}(\mathbf{S},\mathbf{R},\mathbf{R});
\quad
\widetilde{\mathbf{S}}
=
\mathbf{S}
+
\mathbf{H}_r
\odot
\sigma(\mathbf{S}\mathbf{W}_g^r)
\end{equation*}
where $\sigma(\cdot)$ is the logistic sigmoid and $\odot$ denotes element-wise multiplication. 
Here, $\mathbf{W}_g^f$ and $\mathbf{W}_g^r$ are separately learned gating parameters for the forward and reverse directions. This operation contextualizes each sentence with the rows it references. The sentence \ex{$s_1$} describing \ex{\textit{Alice}}'s \ex{\textit{Vancomycin}} prescription attends most strongly to the corresponding patient and medication rows, producing a row-aware representation in $\widetilde{\mathbf{S}}$. Together, the two directions reduce ambiguity by conditioning each modality on the other.

\subsection{Output}
The primary outputs of the LOKI encoder are the contextualized representations $\widetilde{\mathbf{R}}$ and $\widetilde{\mathbf{S}}$, from which LOKI computes two secondary outputs: a fine-grained row-sentence similarity matrix and a coarse-grained table-text similarity score.

\medskip

\stitle{Row-Sentence Similarity Matrix.} The fine-grained association between individual rows and sentences is captured in the matrix $\mathbf{P}\in\mathbb{R}^{n\times m}$, computed using cosine similarity between their contextualized representations, where $r\in\{1,\ldots,n\}$ indexes the table rows and $s\in\{1,\ldots,m\}$ the sentences:
\begin{equation*}
P_{rs}
=
\frac{
\langle\widetilde{\mathbf{R}}_r,\widetilde{\mathbf{S}}_s\rangle
}{
\|\widetilde{\mathbf{R}}_r\|_2
\cdot
\|\widetilde{\mathbf{S}}_s\|_2
}
\end{equation*}
The pair-score matrix preserves the row-sentence granularity required for join-path extraction. In our running example, the \ex{\textit{Alice}} row receives a high score for sentence \ex{$s_1$} describing her prescription and lower scores with sentences concerning unrelated patients.

\medskip

\stitle{Table-Text Similarity} For the global retrieval task and loss calculation, $\mathbf{P}$ is aggregated into a scalar similarity score, $\operatorname{Sim}(T,D)$. Empirically, we adopt \emph{Top-$K$ sum pooling} with $K_{\mathrm{pool}}=5$ by default:
\begin{equation*}
\operatorname{Sim}(T,D)
=
\sum_{(r,s)\in\operatorname{TopK}(\mathbf{P};K_{\mathrm{pool}})}
P_{rs}
\end{equation*}
Here, $\operatorname{TopK}(\mathbf{P};K_{\mathrm{pool}})$ denotes the index pairs $(r,s)$ corresponding to the $K_{\mathrm{pool}}$ largest entries of $\mathbf{P}$. Aggregating only the strongest row-sentence links avoids relying on a single pair, as in max-pooling, prevents dilution by mean-pooling, and concentrates the global ranking gradient on the most relevant local interactions, directly coupling table-text retrieval with fine-grained row-sentence association.

\begin{figure}[t]
  \centering
  \includegraphics[width=1\linewidth]{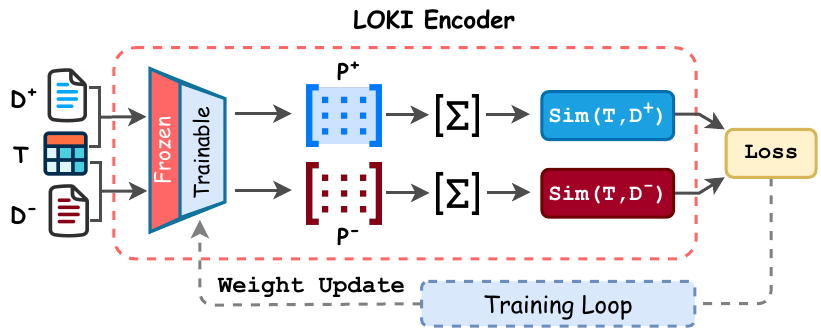}
    \caption{LOKI training pipeline, from coarse table-document supervision to fine-grained row-sentence association.}
  \label{fig:LOKI-Training-Pipeline}
\end{figure}

\subsection{Training}

LOKI's training (Fig.~\ref{fig:LOKI-Training-Pipeline}) uses a composite loss over triplets $(T, D^+, D^-)$, where $D^+$ is a relevant document for table $T$ and $D^-$ is an unrelated or hard-negative document. The total loss jointly optimizes five objectives:
\begin{equation*}
\resizebox{0.99\linewidth}{!}{$
\mathcal{L}_{\mathrm{total}} = \lambda_{\mathrm{glob}}\mathcal{L}_{\mathrm{glob}} + \lambda_{\mathrm{loc}}\mathcal{L}_{\mathrm{loc}} + \lambda_{\mathrm{dist}}\mathcal{L}_{\mathrm{dist}} + \lambda_{\mathrm{sig}}\mathcal{L}_{\mathrm{sig}} + \lambda_{\mathrm{sink}}\mathcal{L}_{\mathrm{sink}}
$}
\end{equation*}
Here, $\mathcal{L}_{\mathrm{glob}}$ drives coarse-grained table-document retrieval from aggregated pair scores, $\mathcal{L}_{\mathrm{loc}}$ sharpens row-sentence separation within the pair-score matrix, and $\mathcal{L}_{\mathrm{dist}}$ aligns LOKI's pair-score distributions with the frozen encoder's. $\mathcal{L}_{\mathrm{sig}}$ and $\mathcal{L}_{\mathrm{sink}}$ regularize the contextualized representations and attention maps, and the $\lambda$ coefficients weight each term.

\medskip

\stitle{Global Table-Text Loss ($\mathcal{L}_{\mathrm{glob}}$)} The global objective encourages the positive table-document pair to receive a higher aggregated similarity than the negative pair:
\begin{equation*}
\mathcal{L}_{\mathrm{glob}}
=
-\log
\frac{\exp(\operatorname{Sim}(T,D^+)/\tau)}
{\exp(\operatorname{Sim}(T,D^+)/\tau)+\exp(\operatorname{Sim}(T,D^-)/\tau)}
\end{equation*}
where $\tau$ is a temperature parameter that controls the sharpness of the contrastive comparison between positive and negative table-document pairs; we used $\tau=0.2$ by default. Since $\operatorname{Sim}(T,D)$ is aggregated from the pair-score matrix, this objective propagates table-text supervision to the strongest row-sentence interactions.

\medskip

\stitle{Local Row-Sentence Loss ($\mathcal{L}_{\mathrm{loc}}$)} To sharpen fine-grained association, we contrast a sentence $s^+$ from the positive document with a sentence $s^-$ from the negative document, and expect the former to have a stronger similarity to a row $r$ compared to the latter. Let $\Omega = \{ r, s^+, s^- \}$ denote the constrasting triples; the set is constructed as follows. We first sample a row $r$ that has the highest similarity with some sentence in $D^+$. Then, we sample a sentence $s^+$ from $D^+$ among those with the highest similarity to $r$. Finally, we sample a sentence $s^-$ from $D^-$. We define the following local objective:
\begin{equation*}
\mathcal{L}_{\mathrm{loc}}
=
\frac{1}{|\Omega|}
\sum_{(r,s^+, s^-)\in\Omega}
\max\Big\{
0,\,
m_{\mathrm{loc}}
-
\big(
P^+_{rs^+}
-
P^-_{rs^-}
\big)
\Big\}
\end{equation*}
Using a local margin of $m_{\mathrm{loc}}=0.3$ by default, the loss encourages row-sentence interactions in the positive context to score higher than those in the negative context by at least this margin. The resulting loss is averaged over the training triplets. This term sharpens the final pair-score matrix \emph{without requiring any ground-truth row-sentence association labels} and complements the global ranking loss by enforcing local discriminability between positive and negative contexts.

\medskip

\stitle{Distillation Loss ($\mathcal{L}_{\mathrm{dist}}$)} Because the global ranking objective uses Top-$K$ aggregation, its supervision is concentrated on the strongest row-sentence pairs. We therefore distill the broader similarity structure of the frozen encoder into LOKI's pair-score matrix $\mathbf{P}$; the teacher scores are simply the cosine similarities between the original non-contextualized embeddings $\mathbf{R}$ and $\mathbf{S}$, requiring no separate teacher model. To reduce the influence of generic sentences that are similar to many rows, we center each teacher score by the mean score received by its sentence:
\begin{equation*}
P^{\xi}_{rs}
=
\frac{\langle \mathbf{R}_r,\mathbf{S}_s\rangle}
{\|\mathbf{R}_r\|\cdot\|\mathbf{S}_s\|};
\qquad
\widehat{P}^{\xi}_{rs}
=
P^{\xi}_{rs}
-
\frac{1}{n}\sum_{r'=1}^{n}P^{\xi}_{r's}
\end{equation*}

We then construct two conditional distributions from the centered teacher scores and the contextualized student scores:
\begin{equation*}
\begin{aligned}
\mathbf{D}^{\xi}_{\mathbf{R}\rightarrow\mathbf{S}}
&=
\operatorname{softmax}_{s}\left(\frac{\widehat{\mathbf{P}}^{\xi}}{\tau_t}\right);
\quad
\mathbf{D}^{\mathrm{LOKI}}_{\mathbf{R}\rightarrow\mathbf{S}}
=
\operatorname{softmax}_{s}\left(\frac{\mathbf{P}}{\tau_s}\right);
\\
\mathbf{D}^{\xi}_{\mathbf{S}\rightarrow\mathbf{R}}
&=
\operatorname{softmax}_{r}\left(\frac{\widehat{\mathbf{P}}^{\xi}}{\tau_t}\right);
\quad
\mathbf{D}^{\mathrm{LOKI}}_{\mathbf{S}\rightarrow\mathbf{R}}
=
\operatorname{softmax}_{r}\left(\frac{\mathbf{P}}{\tau_s}\right)
\end{aligned}
\end{equation*}
Here, $\operatorname{softmax}_{s}$ normalizes over sentences for each row, whereas $\operatorname{softmax}_{r}$ normalizes over rows for each sentence. We use teacher and student temperatures $\tau_t=0.5$ and $\tau_s=0.1$, respectively. LOKI matches the teacher and student distributions in both normalization directions using Jensen-Shannon divergence:
\begin{equation*}
\mathcal{L}_{\mathrm{dist}}
=
\frac{1}{2}
\left[
\operatorname{JS}\!\left(
\mathbf{D}^{\xi}_{\mathbf{R}\rightarrow\mathbf{S}}
\,\|\, 
\mathbf{D}^{\mathrm{LOKI}}_{\mathbf{R}\rightarrow\mathbf{S}}
\right)
+
\operatorname{JS}\!\left(
\mathbf{D}^{\xi}_{\mathbf{S}\rightarrow\mathbf{R}}
\,\|\, 
\mathbf{D}^{\mathrm{LOKI}}_{\mathbf{S}\rightarrow\mathbf{R}}
\right)
\right]
\end{equation*}
The first term preserves each row's relative association across sentences, while the second preserves each sentence's relevance across rows. Computed for both positive and negative contexts and averaged, this loss preserves associations beyond the strongest Top-$K$ pairs without directly supervising the attention matrices $\mathbf{A}_f$, $\mathbf{A}_r$.

\medskip

\stitle{Regularization Losses ($\mathcal{L}_{\mathrm{sig}}$, $\mathcal{L}_{\mathrm{sink}}$)} We use two complementary regularizers to prevent representation and attention collapse. First, the SIGReg regularizer of~\cite{SIGRegLeCun2025} encourages diversity and approximate isotropy in the contextualized embeddings by penalizing deviations of their random projections from a Gaussian distribution:
\begin{equation*}
\mathcal{L}_{\mathrm{sig}}
=
\frac{1}{4}
\Big[
\operatorname{SIGReg}(\widetilde{\mathbf{R}}^+)
+
\operatorname{SIGReg}(\widetilde{\mathbf{S}}^+)
+
\operatorname{SIGReg}(\widetilde{\mathbf{R}}^-)
+
\operatorname{SIGReg}(\widetilde{\mathbf{S}}^-)
\Big]
\end{equation*}

Second, a Sinkhorn-inspired marginal constraint~\cite{SinkhornAude2018} discourages hub-like attention, in which a small number of rows or sentences absorb most of the attention mass:
\begin{equation*}
\begin{aligned}
\operatorname{Sk}(\mathbf{A})
&=
\frac{1}{N_k}
\left\|
\operatorname{col\_sum}(\mathbf{A})
-
\frac{N_q}{N_k}\mathbf{1}
\right\|_2^2
+
\operatorname{Var}\!\left(
\operatorname{col\_sum}(\mathbf{A})
\right);
\\
\mathcal{L}_{\mathrm{sink}}
&=
\frac{1}{4}
\Big[
\operatorname{Sk}(\mathbf{A}^+_f)
+
\operatorname{Sk}(\mathbf{A}^+_r)
+
\operatorname{Sk}(\mathbf{A}^-_f)
+
\operatorname{Sk}(\mathbf{A}^-_r)
\Big]
\end{aligned}
\end{equation*}
Here, $\operatorname{col\_sum}(\mathbf{A})$ denotes the column sums of the row-normalized attention matrix, while $N_q$ and $N_k$ denote the numbers of valid queries and keys, respectively. The target column sum $N_q/N_k$ corresponds to distributing the attention mass uniformly across the keys. Together, these regularizers maintain diverse contextualized representations and balanced bidirectional attention distributions.

\section{Text-Table Integration}
We now discuss how LOKI discovers text-mediated join paths and uses them to integrate tables. Let $\mathcal{T} = \{T_k\}_{k=1}^N$ be the collection of tables in a raw multi-modal data lake and $\mathcal{D} = \{D_\ell\}_{\ell=1}^M$ its collection of documents.

\medskip

\stitle{Step 1: Candidate Discovery.}
We assume that LOKI has been trained and denote the trained model by $\operatorname{LOKI}_{\theta}$, where $\theta$ represents its learned parameters while the sentence encoder $\xi$ remains frozen. We first discover coarsely associated tables and documents directly from the encoder's outputs: for every table--document pair $(T_k,D_\ell)$, $\operatorname{LOKI}_{\theta}$ aggregates its pair-score matrix $\mathbf{P}$ into the global similarity $q_{k\ell}=\operatorname{Sim}(T_k,D_\ell)$. We retain the sufficiently similar pairs as the set $\mathcal{A}$ of candidate table-document associations:
\begin{equation*}
\mathcal{A}
=
\left\{
(T_k,D_\ell)\in\mathcal{T}\times\mathcal{D}
\mid q_{k\ell} \geq \delta_{\mathrm{ret}}
\right\},
\end{equation*}
where the retrieval threshold $\delta_{\mathrm{ret}}$ is selected on the validation set and fixed during inference to bound the number of combinations subsequently considered for join-path extraction. Candidate combinations are then constructed by pairing distinct tables associated with the same document:
\begin{equation*}
\mathcal{C}
=
\left\{
(T_A,D,T_B)
\mid
(T_A,D)\in\mathcal{A},
(T_B,D)\in\mathcal{A}
\right\}
\end{equation*}
Each candidate combination $(T_A,D,T_B)\in\mathcal{C}$ therefore contains two tables independently relevant to $D$; documents with fewer than two retained tables yield no candidate.

\subsection{Text-Mediated Join Path Discovery}

Having constructed $\mathcal{C}$, LOKI performs fine-grained discovery within each candidate combination. For simplicity, let $(T_A,D,T_B)\in\mathcal{C}$ denote an arbitrary candidate, where $T_A$ and $T_B$ are schematically disjoint tables independently identified as relevant to $D$. Their coarse associations with $D$ do not yet establish how their rows are related. For each candidate, $\operatorname{LOKI}_{\theta}$ jointly contextualizes the rows from both tables with the sentences in $D$ while preserving their table of origin: we linearize and encode their rows before concatenating the resulting sequences in a fixed order:
\begin{equation*}
\mathbf{R}_0
=
\xi\big(\operatorname{linearize}(T_A)\big)
\oplus
\xi\big(\operatorname{linearize}(T_B)\big);
\quad
\mathbf{S}_0
=
\xi\big(\operatorname{sentences}(D)\big)
\end{equation*}
where $\mathrm{sentences}(\cdot)$ segments a document into sentences, $\oplus$ denotes ordered concatenation, and $n_A$ and $n_B$ are the row counts of $T_A$ and $T_B$, so that the first $n_A$ rows originate from $T_A$ and the following $n_B$ from $T_B$. A single joint forward pass then produces the contextualized representations and pair-score matrix:
\begin{equation*}
(\widetilde{\mathbf{R}},\widetilde{\mathbf{S}},\mathbf{P})
=
\operatorname{LOKI}_{\theta}(\mathbf{R}_0,\mathbf{S}_0).
\end{equation*}
These outputs provide a shared row-sentence context from which sparse, evidence-backed join paths can be extracted. Algorithm~\ref{algo:inference-loki} summarizes this join-path extraction process.

\medskip

\stitle{Step 2: Atomic Link Identification}
The pair-score matrix $\mathbf{P}$ is dense because every row in the ordered sequence $\mathbf{R}_0$ receives an affinity score against every sentence in $D$. Since only a small subset provides useful evidence for text-mediated join paths, $\operatorname{LOKI}_{\theta}$ converts $\mathbf{P}$ into a sparse set of high-confidence atomic links. For each candidate combination $(T_A,D,T_B)$, we compute an adaptive threshold $\gamma$ rather than using a globally fixed cutoff. For each row $r$, let $q_r=\max_s(P_{rs})$ denote its highest score across all sentences, and let $P_{75}$ denote the 75th percentile of $\{q_r\}_{r=1}^{n_A+n_B}$. Furthermore, $\mu_{\mathbf{P}}$ and $s_{\mathbf{P}}$ denote the mean and standard deviation of all entries in $\mathbf{P}$. The adaptive threshold is given by:
\begin{equation*}
\gamma
=
\max\!\left(
\min\!\left(
P_{75},
\mu_{\mathbf{P}}+2s_{\mathbf{P}}
\right),
\gamma_{\min}
\right)
\end{equation*}
where $\gamma_{\min}=0.15$ is the default minimum score threshold. This makes the extraction stricter when the example contains clear row-sentence associations, while still preventing the threshold from collapsing on noisier examples. We then apply mutual Top-$k$ filtering. For each row $r_i$, the filter retains the indices of its $k_r$ highest-scoring sentences, whereas for each sentence $s_t$, it retains the indices of its $k_s$ highest-scoring rows:
\begin{equation*}
\mathcal{N}^{\mathrm{row}}_{k_r}(i)
=
\operatorname{Top}_{k_r}(P_{i,:});
\quad
\mathcal{N}^{\mathrm{sent}}_{k_s}(t)
=
\operatorname{Top}_{k_s}(P_{:,t})
\end{equation*}
where $\operatorname{Top}_k(\cdot)$ returns the indices of the $k$ largest entries. The sentence-side operation is applied independently to the row partitions of $T_A$ and $T_B$. We use $k_r=32$ and $k_s=10$ by default. The row-side filter prevents a row from attaching to too many vague sentences, while the sentence-side filter suppresses hub sentences that align weakly with many rows but do not provide discriminative evidence for a specific join path. The resulting atomic-link set is:
\begin{equation*}
\mathcal{J}
=
\left\{
(i,t)
\;\middle|\;
P_{it}\geq\gamma
\;\land\;
t\in\mathcal{N}^{\mathrm{row}}_{k_r}(i)
\;\land\;
i\in\mathcal{N}^{\mathrm{sent}}_{k_s}(t)
\right\}
\end{equation*}
Since Step 3 must connect rows across two schematically disjoint tables, we partition these links by table origin:
\begin{equation*}
    \begin{aligned}
    \mathcal{J}_A &= \{(i,t)\in\mathcal{J}:1\le i\le n_A\} \\
    \mathcal{J}_B &= \{(j,t):(n_A+j,t)\in\mathcal{J},\;1\le j\le n_B\}
\end{aligned}
\end{equation*}
Here, $\mathcal{J}_A$ and $\mathcal{J}_B$ preserve table origin while retaining the sentence supporting each atomic link. These table-specific links provide the evidence anchors for cross-table join paths.

\begin{algorithm}[]
\SetAlgoLined
\KwIn{Candidate $(T_A,D,T_B)$, frozen encoder $\xi$, trained model $\operatorname{LOKI}_{\theta}$, threshold floor $\gamma_{\min}$, and Top-$k$ parameters $k_r, k_s$}
\KwOut{Atomic links $\mathcal{J}_A,\mathcal{J}_B$ and candidate join paths $\mathcal{P}_{\mathrm{cand}}$}
\BlankLine

\tcp{\textbf{Joint row-sentence encoding}}
$\mathbf{R}_0 \leftarrow \xi\!\left(\operatorname{linearize}(T_A)\oplus\operatorname{linearize}(T_B) \right)$ \;
$\mathbf{S}_0 \leftarrow \xi(\operatorname{sentences}(D))$ \;
$(\widetilde{\mathbf{R}}, \widetilde{\mathbf{S}}, \mathbf{P}) \leftarrow \operatorname{LOKI}_{\theta}(\mathbf{R}_0,\mathbf{S}_0)$ \;

\tcp{\textbf{Compute adaptive score floor}}
$\gamma \leftarrow
\max(\min(P_{75}, \mu_{\mathbf{P}} + 2s_{\mathbf{P}}), \gamma_{\min})$ \;

\tcp{\textbf{Extract table-specific atomic links}}
$\mathcal{N}^{\mathrm{row}}_{k_r}(i)\leftarrow \operatorname{Top}_{k_r}(P_{i,:}), \quad
\mathcal{N}^{\mathrm{sent}}_{k_s}(t)\leftarrow \operatorname{Top}_{k_s}(P_{:,t})$ \;

$\mathcal{J}\leftarrow
\{(i,t)\mid
P_{it}\ge\gamma
\land t\in\mathcal{N}^{\mathrm{row}}_{k_r}(i)
\land i\in\mathcal{N}^{\mathrm{sent}}_{k_s}(t) \}$ \;

$\mathcal{J}_A\leftarrow\{(i,t)\in\mathcal{J}:1\le i\le n_A\}$ \;
$\mathcal{J}_B\leftarrow\{(j,t):(n_A+j,t)\in\mathcal{J},\;1\le j\le n_B\}$ \;

\tcp{\textbf{Construct shared-sentence join paths}}
$\mathcal{P}_{\mathrm{cand}}\leftarrow
\left\{
(r_i^A,s_t,r_j^B)
\;\middle|\;
(i,t)\in\mathcal{J}_A
\land
(j,t)\in\mathcal{J}_B
\right\}$ \;

\ForEach{sentence and each row pair}{retain only the path with maximal $w_{ijt}$}
\Return{$\mathcal{J}_A,\mathcal{J}_B,\mathcal{P}_{\mathrm{cand}}$}
\caption{Join-path extraction (Steps 2--3) for one candidate combination $(T_A, D, T_B)$}
\label{algo:inference-loki}
\end{algorithm}

\stitle{Step 3: Transitive Join-Path Construction}
Because $T_A$ and $T_B$ are schematically disjoint, LOKI does not assume that a row from one table can be joined directly with a row from the other table. Instead, the connection must be mediated by text. After atomic-link identification, a row $r_i^A\in T_A$ and a row $r_j^B\in T_B$ become join candidates only when they share at least one mediating sentence $s_t$. For each shared sentence, LOKI constructs a text-mediated join path and assigns it a bridge weight:
\begin{equation*}
p_{ijt} = (r_i^A,s_t,r_j^B), \quad w_{ijt} = \frac{ P_{it}+P_{n_A+j,t} }{2}
\end{equation*}
Here, $p_{ijt}$ is a single-evidence instance of the general text-mediated join path $p=(r_i^A,S_p,r_j^B)$, with $S_{p_{ijt}}=\{s_t\}$. Paths connecting the same row pair and relationship type are consolidated by collecting their mediating sentences into $S_p$ during materialization. For simplicity, each path in our running example contains a single sentence. Because a candidate path is formed only when $s_t$ has retained atomic links to both rows, $w_{ijt}$ summarizes their joint compatibility and serves as the bridge confidence used to rank candidate paths and weight the contextualized sentence representation during materialization. The candidate path set is then:
\begin{equation*}
\mathcal{P}_{\mathrm{cand}} = \left\{ p_{ijt} = (r_i^A,s_t,r_j^B) \;\middle|\; (i,t)\in\mathcal{J}_A \wedge (j,t)\in\mathcal{J}_B\right\}
\end{equation*}
To avoid over-representing repeated or generic evidence, LOKI retains only the strongest paths per sentence and per row pair. The resulting $\mathcal{P}_{\mathrm{cand}}$ is therefore a compact set of explicit, evidence-backed bridges. In the running example, the sentences \ex{$\{s_1,\ldots,s_5\}$} produce the following paths: \ex{\textit{$p_1=(I_1,\{s_1\in D_1\},M_1)$}}, \ex{\textit{$p_2=(I_1,\{s_2\in D_1\},M_1)$}}, \ex{\textit{$p_3=(I_1,\{s_3\in D_1\},M_1)$}}, \ex{\textit{$p_4=(I_1,\{s_4\in D_1\},M_2)$}}, and \ex{\textit{$p_5=(I_1,\{s_5\in D_1\},M_3)$}}.

\begin{figure*}[]
  \centering
  \includegraphics[width=\linewidth]{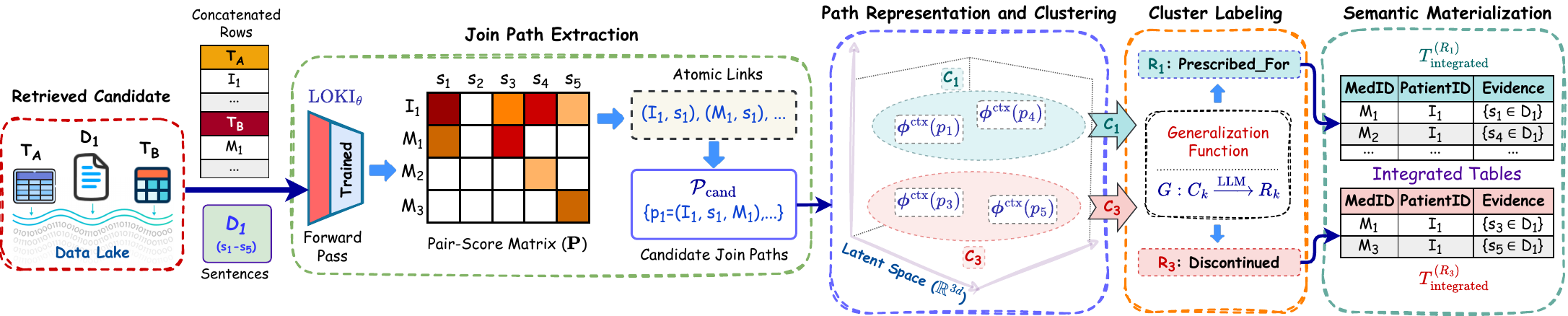}
    \caption{LOKI inference pipeline. (a) Given a retrieved candidate $(T_A,D,T_B)$, a single joint forward pass through $\operatorname{LOKI}_{\theta}$ produces the pair-score matrix $\mathbf{P}$, whose high-confidence atomic links are composed into candidate join paths, (b) clustered into the illustrative groups $\{C_1,C_3\}$. (c) The Generalization Function $G$ assigns the relationship labels $\{R_1,R_3\}$, (d) and the labeled clusters are materialized as relation-specific integrated tables with supporting evidence.}
  \label{fig:example_integration}
\end{figure*}

\subsection{Integration}\label{sec:materialization}


Discovery-driven integration requires mapping candidate join paths to latent relationship types $\mathcal{R}$ and materializing them as typed integrated tables. Each candidate path $p_{ijt}=(r_i^A,s_t,r_j^B)\in\mathcal{P}_{\mathrm{cand}}$ is treated as a distinct evidence unit, allowing different sentences connecting the same row pair to express different relationships. In the running example, \ex{$p_1=(I_1,\{s_1\in D_1\},M_1)$} and \ex{$p_2=(I_1,\{s_2\in D_1\},M_1)$} share the same row pair but express \ex{\textit{Prescribed\_For}} and \ex{\textit{Adverse\_Effect}}, respectively.

\medskip

\stitle{Step 4: Path Representation and Clustering}
To group candidate paths by the relationship expressed in their mediating evidence, LOKI constructs a path-specific representation that combines the contextualized sentence with the strength of its attachment to both table rows. For a candidate path $p_{ijt}=(r_i^A,s_t,r_j^B)$, let
\begin{equation*}
    \alpha_{it}=P_{it},
    \quad
    \beta_{jt}=P_{n_A+j,t},
    \quad
    w_{ijt}=\frac{\alpha_{it}+\beta_{jt}}{2}
\end{equation*}
Here, $\alpha_{it}$ and $\beta_{jt}$ measure how strongly sentence $s_t$ supports rows $r_i^A$ and $r_j^B$, respectively, and $w_{ijt}$ is their average, representing the confidence of the complete bridge. We $L_2$-normalize the contextualized embeddings so that the three components begin on the same scale and, since an unweighted concatenation would treat them as equally important even when one underlying link is considerably weaker, we use the association scores as multiplicative gates:
\begin{equation*}
\boldsymbol{\phi}^{\mathrm{ctx}}(p_{ijt})
=
\left[
\alpha_{it}\widetilde{\mathbf{r}}_i^A
\;\Vert\;
w_{ijt}\widetilde{\mathbf{s}}_t
\;\Vert\;
\beta_{jt}\widetilde{\mathbf{r}}_j^B
\right]
\in\mathbb{R}^{3d}
\end{equation*}
The ordered concatenation preserves the distinct roles of the two table rows and the mediating sentence, while the weights modulate each component according to the strength of its supporting evidence, preserving potentially asymmetric attachment strengths. Thus, paths involving the same sentence remain distinguishable when they connect different row pairs or exhibit different attachment strengths. Finally, the complete path vector is $L_2$-normalized before clustering, preventing its overall magnitude from dominating the clustering distance:
\begin{equation*}
    \mathcal{K}
    =
    \{C_1,\ldots,C_M\}
    =
    \operatorname{HDBSCAN}
    \left(
    \left\{
    \boldsymbol{\phi}^{\mathrm{ctx}}(p)
    \mid
    p\in\mathcal{P}_{\mathrm{cand}}
    \right\}
    \right)
\end{equation*}

Each cluster $C_k \in \mathcal{K}$ therefore groups text-mediated join paths whose row-conditioned evidence expresses similar relational semantics. In the running example, the paths \ex{$p_1$} and \ex{$p_4$} are expected to be grouped together because they represent a \ex{\textit{Prescribed\_For}} relationship, while \ex{$p_3$} and \ex{$p_5$} form another group because they represent a \ex{\textit{Discontinued}} relationship.

\medskip

\stitle{Step 5: Relationship Materialization}
After clustering, each cluster $C_k$ contains text-mediated join paths whose row-conditioned evidence expresses a coherent but unnamed relationship. To make this relationship interpretable, we use a \textit{Generalization Function} $G$ that employs an LLM to summarize the shared evidence within the cluster and assign a descriptive relationship label $R_k$:
\begin{equation*}
    G:C_k\xrightarrow{\mathrm{LLM}}R_k
\end{equation*}
Each labeled cluster is then materialized as an integrated table:
\begin{equation*}
    T_{\mathrm{integrated}}^{(R_k)}
    =
    \left\{
    (r_i^A,r_j^B,s_t)
    \;\middle|\;
    p_{ijt}=(r_i^A,s_t,r_j^B)\in C_k
    \right\}
\end{equation*}

Repeating this process for each labeled cluster produces the final set of relation-specific integrated tables, where every entry combines one record from $T_A$ with one record from $T_B$ and preserves the supporting evidence for the relationship $R_k$. 
Fig.~\ref{fig:example_integration} illustrates the materialization of a subset of the paths in our running example, first represented in the latent space and clustered into relation-consistent groups. The paths \ex{\textit{$p_1$}} and \ex{\textit{$p_3$}} remain close because they share the same row pair \ex{\textit{$(I_1,M_1)$}}, but they are assigned to different clusters because they represent different relationships. The Generalization Function then labels $C_1$ as \ex{\textit{$R_1:\text{Prescribed\_For}$}} and $C_3$ as \ex{\textit{$R_3:\text{Discontinued}$}}. Finally, each labeled cluster is materialized as a separate integrated table with sentence-level provenance.

\section{Experimental Evaluation}

We empirically evaluate LOKI through three experiments that examine successive stages of the same discovery-driven integration pipeline: \textbf{(i)} architectural evaluation of representation learning for implicit association, \textbf{(ii)} multi-modal data discovery, and \textbf{(iii)} data integration. Together, the three experiments establish the progression from association to discovery towards integration; Table~\ref{tab:exp_progression} summarizes this evaluation strategy, and each experiment details its objective in the corresponding subsection. All datasets, scripts, and experimental results are publicly available in our repository.\footnote{\texttt{\textbf{Artifact:}} \url{https://github.com/dtim-upc/LOKI}}

\begin{table}[h]
\centering
\caption{Progressive evaluation of LOKI.}
\label{tab:exp_progression}
\resizebox{\columnwidth}{!}{%
\begin{tabular}{lccc}
\toprule
\textbf{Exp.} & \textbf{Capability} & \textbf{Output} & \textbf{Validates} \\
\midrule
Exp.~1 & Association & Row $\leftrightarrow$ Sentence & Representation Learning\\
Exp.~2 & Discovery & Table $\leftrightarrow$ Document & Cross-modal Retrieval\\
Exp.~3 & Integration & Integrated Relation & Discovery-driven Integration\\
\bottomrule
\end{tabular}
}
\end{table}

\subsection{Datasets}\label{sec:eval_datasets}

No publicly available benchmark natively provides the typed row-sentence-row annotations required by our \textit{Text-Mediated Join Path Discovery} task; existing benchmarks target generation, question answering, or coarse discovery instead. We therefore constructed ground truth over four complementary datasets, summarized in Table~\ref{tab:datasets} and used throughout the three experiments; all of them are publicly released.\footnote{\texttt{\textbf{Datasets:}} \url{https://github.com/dtim-upc/LOKI/tree/main/Datasets}} \textbf{FEVEROUS}~\cite{Aly21Feverous} is an open-domain Wikipedia fact-verification benchmark whose evidence combines table cells with sentences, from which we derive row-sentence associations. We manually annotated \textbf{ProTrix}~\cite{ProTrixZirui2024} using the original question-answer pairs. Although \textbf{Pharma}~\cite{EltabakhKEA23} primarily targets table-text discovery, we repurposed it for row-sentence silver annotation using term frequency--inverse document frequency (TF-IDF) similarity and keyword matching.

Among the datasets we consider, \textbf{MIMIC-IV}~\cite{PhysioNet-mimiciv-3.1,johnson2023mimic} provides the most comprehensive granular links between rows of the two schematically disjoint tables \textit{Diagnosis} and \textit{Medications} through \textit{Clinical Notes}; it is the only dataset that supports our complete end-to-end integration setting. We constructed a validated benchmark from it covering 382 hospital admissions: three frontier LLM annotators independently generated candidate annotations, consolidated through majority voting and verified by a human annotator. All fine-grained annotations are used exclusively for evaluation; the full re-construction procedure, annotation guidelines, and validation protocol are included in the released artifact.

\begin{table}[h]
    \centering
    \caption{Datasets and task coverage. Size denotes total fine-grained (F) and coarse-grained (C) examples.}
    \label{tab:datasets}
    \small
    \setlength{\tabcolsep}{2.0pt}
    \renewcommand{\arraystretch}{1.10}
    \begin{tabular}{@{}p{0.25\columnwidth}p{0.14\columnwidth}p{0.26\columnwidth}p{0.18\columnwidth}p{0.08\columnwidth}@{}}
        \toprule
        \textbf{Dataset} & \textbf{Domain} & \textbf{Native unit} & \textbf{Size (F/C)} & \textbf{Exp.} \\
        \midrule
        \textbf{FEVEROUS}\cite{Aly21Feverous} & Wiki & Cell-Sentence & 500/13,304 & 1 \\
        \textbf{ProTrix}\cite{ProTrixZirui2024} & QA & Table-Paragraph & 446/3,157 & 1 \\
        \textbf{Pharma}\cite{EltabakhKEA23} & BioMed & Tables(8)-Text & 140/926 & 1, 2 \\
        \textbf{MIMIC-IV}\cite{PhysioNet-mimiciv-3.1} & Clinical & Tables(2)-Notes & 382/24,410 & 1, 2, 3 \\
        \bottomrule
    \end{tabular}
\end{table}

\subsection{Experiment 1: Architectural Evaluation}
\label{sec:exp-architecture}

The first experiment investigates whether a model trained exclusively using coarse table-text associations can simultaneously learn: \textbf{(i)} global table-text relevance, required for multi-modal discovery, and \textbf{(ii)} fine-grained row-sentence associations, required for join path extraction. This setting presents a fundamental supervision dilemma. Coarse table-document associations are comparatively easy to obtain, whereas explicitly annotating every supporting row-sentence pair is expensive and domain-dependent, and such annotations are scarce. Standard sentence encoders represent rows and sentences independently and provide no mechanism for identifying which local interactions explain the relevance of an entire table-document pair. LOKI instead derives the global score from a pair-score matrix over mutually contextualized representations, so we can ask whether local association emerges from the global objective alone.

\medskip

\stitle{Investigated Architectures} We compare five configurations. The \textbf{Baseline} uses the \textit{Frozen Encoder} $\xi$ alone to independently embed linearized rows and document sentences, scoring each row-sentence pair by cosine similarity without task-specific training or cross-attention. \textbf{FT-Encoder} fine-tunes $\xi$ using the global table-text triplet objective. \textbf{Uni (R$\rightarrow$S)} contextualizes rows with sentences, whereas \textbf{Uni (S$\rightarrow$R)} contextualizes sentences with table rows, while keeping $\xi$ frozen. \textbf{LOKI} uses the complete bidirectional architecture to jointly contextualize both modalities before constructing the pair-score matrix and aggregated global score (Fig.~\ref{fig:pipeline_loki}).

\medskip

\stitle{Evaluation Protocol} We evaluate all five architectures on all datasets of Table~\ref{tab:datasets}. The fine-grained row-sentence annotations are used only for evaluation. We report the best validation accuracy (\emph{Acc.}) for global table-text matching and track test average precision (\emph{AP}) and macro-$F1$ across epochs for local row-sentence association. AP is the primary local metric under class imbalance, while macro-$F1$ captures the precision-recall balance. We additionally report ranking quality, training time, and peak VRAM/RAM consumption.

\medskip

\stitle{Cross-dataset Architectural Comparison} Table~\ref{tab:exp1-cross-dataset} shows that LOKI's architecture generalizes across benchmarks and transfers across granularities, attaining the best local AP on every dataset and demonstrating that coarse supervision can induce stronger local association. The smaller local gain on Pharma reflects its silver annotations, derived from TF-IDF and keyword matching, which favor explicit lexical matches already captured by the Baseline and provide limited scope for measuring deeper semantic association. Overall, bidirectional contextualization transfers table-text supervision to row-sentence association more effectively than encoder fine-tuning or unidirectional attention; next we examine this capability in detail.

\begin{table}[h]
\centering
\caption{Table-Text accuracy (\emph{Acc.}) vs. row-sentence average precision (\emph{AP}) across datasets, measuring implicit learning from the coarse training task to fine-grained association.}
\label{tab:exp1-cross-dataset}
\resizebox{\columnwidth}{!}{%
\begin{tabular}{lcccc}
\hline
\textbf{Configuration}   & \textbf{FEVEROUS} & \textbf{ProTrix} & \textbf{Pharma}  & \textbf{MIMIC-IV}      \\ \hline
Baseline         &  $0.96$ / $0.34$              & $0.86$ / $0.17$          & $0.91$ / $0.40$          & $0.80 / 0.42$          \\
FT-Encoder             &  $0.96$ / $0.34$              & $0.93$ / $0.58$          & $0.93$ / $0.40$          & $0.81 / 0.44$          \\
Uni (R$\rightarrow$S)  &  $0.77$ / $0.35$              & $0.83$ / $0.62$          & $1.0$ / $0.40$          & $0.73$ / $0.42$                \\
Uni (S$\rightarrow$R)  &  $0.82$ / $0.33$              & $0.55$ / $0.39$          & $1.0$ / $0.39$          & $0.72$ / $0.41$                \\
\textbf{LOKI}          &  $\mathbf{0.98 / 0.44}$     & $\mathbf{0.94 / 0.67}$  & $\mathbf{0.99 / 0.42}$ & $\mathbf{0.94 / 0.54}$ \\ \hline
\end{tabular}%
}
\end{table}

MIMIC-IV is more challenging than the other benchmarks: its long, noisy clinical notes express relationships only implicitly, beyond lexical overlap, and its validated fine-grained annotations support the complete end-to-end integration task. It is also where the architectural differences become clearest. Although the Baseline uses a domain-specific encoder (MedEmbed),\footnote{\texttt{\textbf{Baseline encoder:}}~\url{https://github.com/abhinand5/MedEmbed}} fine-tuning it transfers poorly from the global objective to local association, and the unidirectional variants retain near-Baseline AP while degrading global accuracy. LOKI outperforms all of them on both granularities.

\begin{figure}[h]
    \centering
    \includegraphics[width=1.0\linewidth]{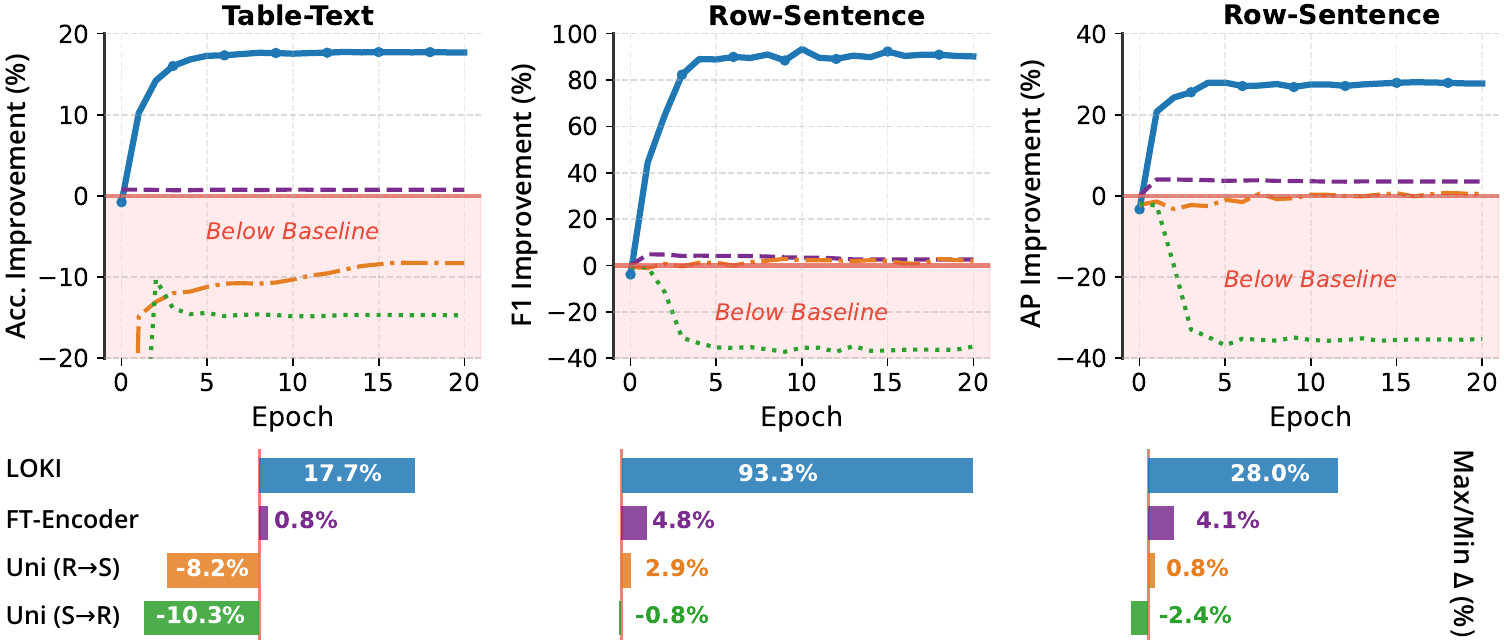}
    \caption{Relative improvement over the Baseline on global Acc. and local macro-$F1$ and AP across training epochs.}
    \label{fig:loki-vs-baseline}
\end{figure}

\stitle{Architectural Benefits} Fig.~\ref{fig:loki-vs-baseline} tracks whether optimization of the coarse table-text task induces fine-grained association, reporting the relative improvement over the Baseline, $\Delta_{\mathrm{rel}}(M)=(M_{\mathrm{model}}-M_{\mathrm{baseline}})/M_{\mathrm{baseline}}\times100\%$, per metric $M$. FT-Encoder produces only small local gains, while the unidirectional variants show uneven transfer; LOKI improves all three metrics, nearly doubling macro-$F1$. All configurations converge within a few epochs, reflecting their strong pre-trained initialization rather than low task complexity: the variants plateau at clearly different levels, and only LOKI's bidirectional contextualization converges to substantially stronger local association using only table-text signals.

\begin{figure}[h]
    \centering
    \includegraphics[width=0.50\columnwidth]{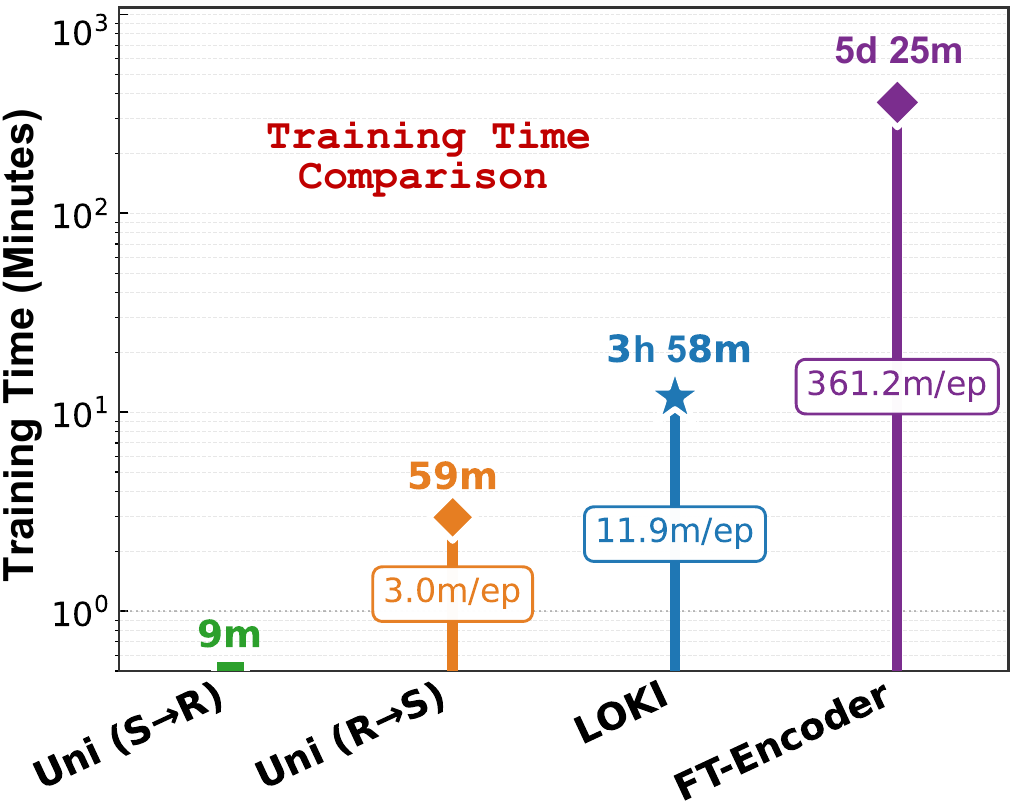}
    \hfill
    \includegraphics[width=0.49\columnwidth]{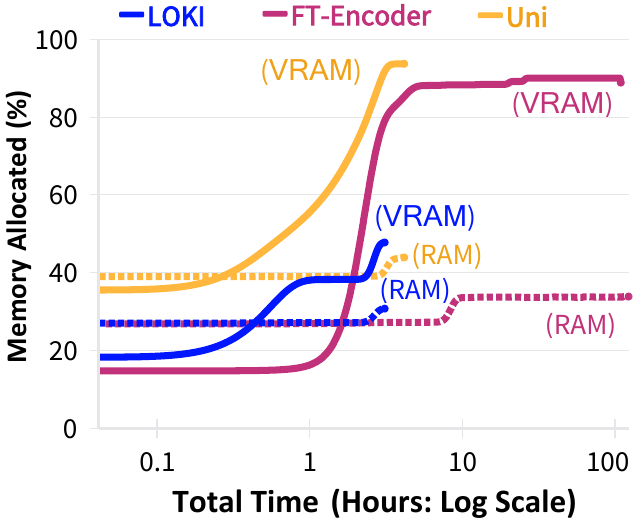}
    \caption{Training time and VRAM/Memory consumption.}
    \label{fig:resource-util}
\end{figure}

\stitle{Compute Requirements} Fig.~\ref{fig:resource-util} shows that LOKI trains approximately $30\times$ faster than full encoder fine-tuning while achieving higher AP/$F1$. On a consumer-grade GPU with 16 GB of VRAM, it also has the lowest peak VRAM and RAM footprint among the trained variants, yielding the best effectiveness-efficiency trade-off.

\medskip

\begin{figure*}[t]
    \centering
    \includegraphics[width=1.0\linewidth]{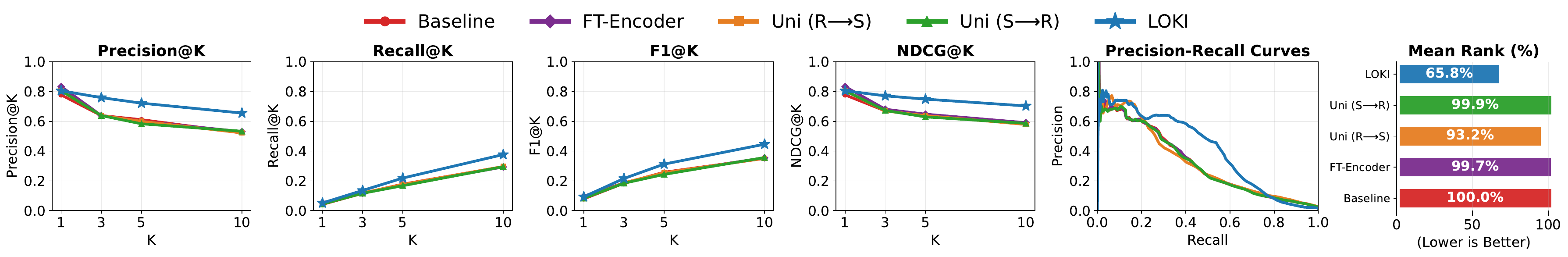}
    \caption{MIMIC-IV row-sentence ranking across cutoff-based metrics, precision-recall, and mean rank.}
    \label{fig:example-ranking-metrics}
\end{figure*}

\stitle{Ranking Analysis} Fig.~\ref{fig:example-ranking-metrics} confirms that LOKI's AP gain reflects consistently stronger rankings. LOKI achieves the highest P@$K$, F1@$K$, and normalized discounted cumulative gain (NDCG@$K$) across all evaluated values of $K$ and maintains the strongest precision-recall curve. Its normalized mean rank is approximately $34\%$ lower than the Baseline, denoting valid evidence appears earlier in the ranking.

\medskip

\stitle{Summary} Exp.~1 shows that off-the-shelf encoder fine-tuning and unidirectional attention do not reliably transfer coarse table-text supervision to fine-grained row-sentence association. LOKI finds better associations in both granularities without local supervision, while remaining more efficient than full encoder fine-tuning.


\subsection{Experiment 2: Multi-Modal Data Discovery}

The second experiment evaluates whether LOKI's global table-text score, derived from fine-grained row-sentence associations, supports accurate and scalable multi-modal data discovery. \textbf{First}, we evaluate whether LOKI learns both retrieval directions: unlike existing systems that are typically designed for a fixed query direction, LOKI learns a shared bidirectional representation, and we test whether a single trained instance can serve both \textit{document-to-table} and \textit{table-to-document} retrieval without retraining. \textbf{Second}, we compare it against state-of-the-art systems operating at the coarser table-document level and examine whether its bottom-up aggregation remains effective as the candidate pool grows.

\medskip

\stitle{Compared Systems} No existing system targets our fine-grained row-sentence discovery task, but several state-of-the-art approaches operate at the coarser table-document level. We therefore evaluate whether LOKI identifies the same information as these systems on their own coarse-grained task, while additionally providing the row-sentence traceability that enables the integration task of Exp.~3. We compare against three representative systems. \textbf{CMDL}~\cite{EltabakhKEA23} targets cross-modal data discovery and was primarily evaluated on the \textbf{Pharma} benchmark for document-to-table retrieval, motivating our use of the same dataset. \textbf{TaBERT}~\cite{YinNYR20} represents joint table-text representation learning, while \textbf{TabSTAR}~\cite{tabSstar2025} represents recent text-aware tabular foundation models. 

\medskip

\stitle{Pharma Benchmark Construction} To ensure a fair comparison with CMDL~\cite{EltabakhKEA23}, we reconstruct Pharma from its $926$ PubMed abstracts and $82$ DrugBank tables. In the original dataset, $8$ tables, spanning $15$ ground-truth columns, are positively associated with every document, while the remaining $74$ tables form the negative pool. We horizontally partition the source tables into row-disjoint candidate tables, ensuring that training, validation, and test splits do not reuse the same rows. For each query, the associated tables form the positive candidates and are paired with an equal number of randomly sampled negative candidates. We additionally inject non-matching rows into positive candidates to prevent models from relying only on isolated lexical matches. The test split contains $140$ query documents and $2{,}240$ candidate tables in the \textit{Full} retrieval pool. This construction requires models to infer query-specific table relevance rather than memorize stable table or column associations.

\medskip

\stitle{Evaluation Protocol} Following the objective, we first evaluate whether LOKI learns both directions: we report cross-dataset \textit{Mean Average Precision} (\textbf{MAP}) on \textbf{Pharma} and \textbf{MIMIC-IV} for \textit{document-to-table} and \textit{table-to-document} retrieval, and then evaluate cross-direction generalization on Pharma. The architecture is direction-agnostic; only the training triplets are anchored on one modality. We call a configuration \emph{direction-matched} when the training anchor and the retrieval query use the same modality, and \emph{direction-mismatched} otherwise; the latter tests whether a single trained instance can serve either query modality without retraining. These comparisons use a fixed pool of 50 candidates per query to control retrieval difficulty. Second, following \textbf{CMDL}'s native \textit{document-to-table} setting, we progressively increase the candidate-table pool and report F1@$K$, NDCG@$K$, mean reciprocal rank (MRR@$K$), MAP, mean rank, and inference time to assess retrieval quality and scalability under realistic data-lake conditions.

\begin{table}[h]
\centering
    \caption{LOKI's cross-dataset table-text discovery (MAP).}
    \label{tab:exp2-cross-dataset}
    \small
    \setlength{\tabcolsep}{2.0pt}
    \renewcommand{\arraystretch}{1.10}
    \begin{tabular}{lcc}
    \hline
    \textbf{Dataset} & \textbf{Doc$\rightarrow$Table} & \textbf{Table$\rightarrow$Doc} \\ \hline
    \textbf{Pharma}    & $0.78$ & $0.20$ \\
    \textbf{MIMIC-IV}  & $0.42$ & $0.54$ \\ \hline
    \end{tabular}
\end{table}

\stitle{Cross-dataset Bidirectional Discovery} On both datasets, LOKI supports both retrieval directions without retraining (Table~\ref{tab:exp2-cross-dataset}), with direction-dependent behavior reflecting the distinct discovery workflows of each benchmark. On MIMIC-IV, it reaches 0.42 MAP for document-to-table and 0.54 for table-to-document retrieval, whereas the latter is noticeably weaker on Pharma. This follows from Pharma's construction: every document is associated with the same eight source tables, so a table query faces a large, diffuse set of relevant abstracts with limited discriminative signal. Existing systems are typically specialized for one query direction and require separate training or retrieval pipelines for the reverse task. LOKI instead provides a single representation that supports both query modalities without retraining.

\begin{table}[h]
\centering
\caption{Cross-direction generalization on Pharma (MAP), crossing the training anchor modality with the retrieval query modality. Grey cells mark direction-mismatched training configurations for retrieval.}\label{tab:exp2-cross-direction}
\label{tab:exp2-cross-direction}
\resizebox{\columnwidth}{!}{%
\begin{tabular}{lcclcc}
\toprule
\multirow{2}{*}{\textbf{System}} & \multicolumn{2}{c}{\textbf{Doc$\rightarrow$Table}} &  & \multicolumn{2}{c}{\textbf{Table$\rightarrow$Doc}} \\ \cline{2-3} \cline{5-6} 
                                 & \textbf{Doc-trained}    & \textbf{Table-trained}   &  & \textbf{Doc-trained}    & \textbf{Table-trained}   \\ \midrule
CMDL          & $0.25$          & \cellcolor{gray!20}$0.20$          &  & \cellcolor{gray!20}$0.09$          & $0.09$          \\
TaBERT        & $0.52$          & \cellcolor{gray!20}$0.23$          &  & \cellcolor{gray!20}$0.11$          & $0.11$          \\
TabSTAR       & $0.20$          & \cellcolor{gray!20}$0.20$          &  & \cellcolor{gray!20}$0.09$          & $0.09$          \\
\textbf{LOKI} & $\mathbf{0.78}$ & \cellcolor{gray!20}$\mathbf{0.35}$ &  & \cellcolor{gray!20}$\mathbf{0.18}$ & $\mathbf{0.20}$ \\ \bottomrule
\end{tabular}%
}
\end{table}

\medskip

\stitle{Cross-Direction Generalization under Direction Mismatch} Table~\ref{tab:exp2-cross-direction} crosses the modality used as the training anchor with the modality used as the retrieval query on Pharma dataset. LOKI achieves the highest MAP in all four settings. Under direction mismatch (grey cells), it remains more than 50\% above the strongest baselines, although document-to-table retrieval is more affected by the mismatch. The main result is therefore not direction-invariant performance, but that a single trained LOKI instance can support either retrieval direction without retraining.


\begin{figure}[h]
  \centering
  \includegraphics[width=\columnwidth]{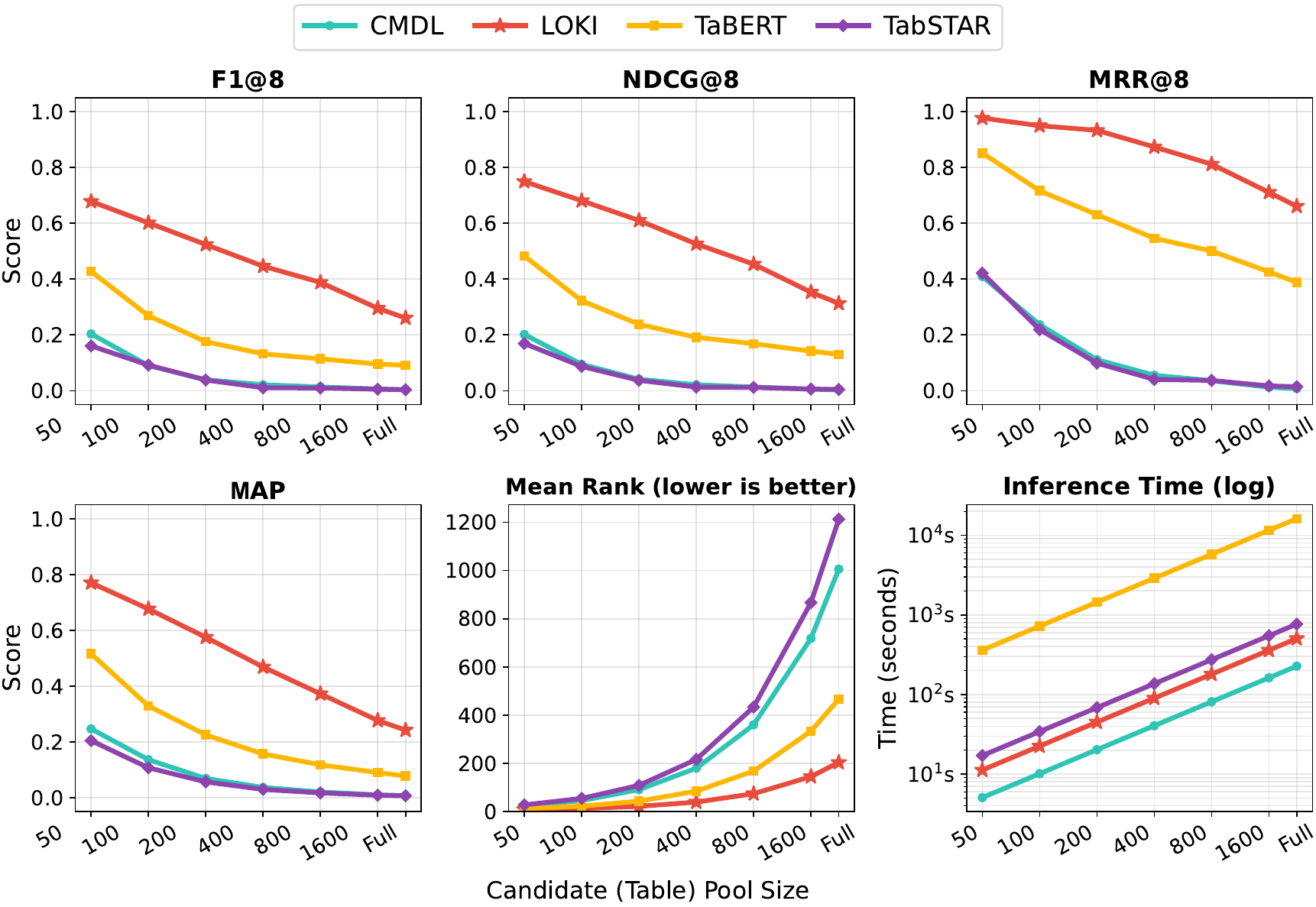}
  \caption{Impact of scaling on data discovery; candidate pool size denotes the number of tables per query document.}
  \label{fig:data_discovery_scalability}
\end{figure}

\stitle{Scalability Analysis} Fig.~\ref{fig:data_discovery_scalability} evaluates \textit{document-to-table} retrieval as the candidate-table pool increases from $50$ to the \textit{Full} set of $2{,}240$ candidates. We report metrics at $K=8$ because each Pharma query document is associated with eight relevant tables. LOKI consistently achieves the strongest retrieval quality across all pool sizes and degrades less drastically than the competing systems. At the \textit{Full} scale it roughly triples the F1@8 and MAP of the second-strongest TaBERT, achieves the lowest mean rank by placing relevant tables approximately 56\% earlier, and completes full-pool retrieval $32\times$ faster than TaBERT. Although CMDL is faster still due to its ANN-based approximate retrieval, its retrieval quality collapses at scale, with MAP approximately $35\times$ lower than LOKI's.

\medskip

\stitle{Summary} Exp.~2 demonstrates that LOKI's fine-grained row-sentence associations provide a strong basis for multi-modal discovery: a single trained instance serves either retrieval direction without retraining, and remains the strongest method as the candidate pool grows.



\subsection{Experiment 3: Data Integration}
\label{sec:exp3_integration}

The third experiment evaluates whether fine-grained row-sentence associations underlying table-document discovery can be transformed into explicit \emph{text-mediated join paths} and subsequently materialized as relationship-specific \emph{Typed Tables}, completing the progression from \emph{association} to \emph{discovery} and finally \emph{integration}.

\medskip

\stitle{Evaluation Protocol}
We evaluate end-to-end integration on 382 MIMIC-IV admissions, each constituting an independent integration problem over the schematically disjoint \textit{Diagnosis} and \textit{Medications} tables and their associated \textit{Clinical Notes}. For each admission, LOKI transforms the discovered text-mediated join paths into provenance-preserving Typed Tables. Since the Generalization Function $G$ produces open-world relationship names, we ground them to the canonical relation types $\mathcal{R}$ before comparing the predicted Typed Tables with their ground-truth counterparts.

\medskip

\stitle{Compared Systems}
We instantiate $G$ with two interchangeable open-weight local LLMs, GPT-OSS 20B and Qwen-3.6, denoted as \textbf{LOKI (GPT-OSS)} and \textbf{LOKI (Qwen-3.6)}, respectively. We compare LOKI against direct LLM-based integration using the same local \textbf{Qwen-3.6}, and additionally include \textbf{Qwen-3.7} as a stronger API-based frontier baseline. The direct LLMs receive the relevant Diagnosis table, Medications table, and Clinical Notes directly and infer typed row-pair relationships for materialization as Typed Tables under the same canonical schema, thereby evaluating relation inference over known inputs rather than table-document discovery or text-mediated join-path recovery.

\medskip

\stitle{Evaluation Metrics}
We evaluate the outputs at two complementary levels. At the \emph{typed-pair level}, \textit{Best-Match Typed-Pair} precision (P), recall (R), and F1 measure row-pair correctness and coverage after each predicted Typed Table is associated with its best-matching ground-truth relationship type. The reported summary scores are macro-averaged over admissions. At the \emph{Typed Table level}, \textit{Type Acc.} measures whether the predicted table is assigned the correct canonical relationship type, while Typed Table P, R, and F1 measure the final relationship-specific materialization quality and are likewise macro-averaged over admissions. For both levels we additionally report pooled counts and the corresponding micro scores, computed on the same evaluation surface as the macro metrics. For efficiency we report average runtime per admission, LLM token consumption, and estimated API-equivalent cost over the 382-admission test set.

\medskip

\noindent We organize the evaluation around two questions:
\begin{itemize}[label=, leftmargin=2mm]
\setlength\itemsep{1mm}
    \item \textbf{RQ1:} How accurately are relationship semantics disambiguated and materialized as Typed Tables?
    \item \textbf{RQ2:} How practical is discovery-driven integration?
\end{itemize}

\medskip

\stitle{RQ1: How accurately are relationship semantics disambiguated and materialized as Typed Tables?}\\
Recovering a row pair alone is insufficient for semantic integration because the relationship expressed between the participating rows must also be distinguished. We therefore evaluate disambiguation through its final database-level consequence: whether the discovered relationships are assigned to the correct relationship-specific Typed Tables and whether their row-pair contents are faithfully materialized. Table~\ref{tab:exp3_typed_summary} summarizes the main results.

\begin{table}[h]
    \centering
    \caption{Comparison of LOKI and direct LLM baselines on relationship disambiguation and Typed Table materialization. P, R, and F1 are macro-averaged over admissions.}
    \label{tab:exp3_typed_summary}
    \resizebox{\columnwidth}{!}{%
    \begin{tabular}{lccc|cccc}
        \toprule
        & \multicolumn{3}{c|}{\textbf{Best-Match Typed-Pair}}
        & \multicolumn{4}{c}{\textbf{Typed Table Materialization}} \\
        System & P & R & F1 & Type Acc. & P & R & F1 \\
        \midrule

        \textbf{LOKI} (GPT-OSS)
        & 0.982 & 0.486 & 0.627
        & 0.840 & 0.755 & 0.755 & 0.742 \\

        \textbf{LOKI} (Qwen-3.6)
        & 0.982 & 0.515 & 0.652
        & 0.807 & 0.750 & 0.732 & 0.726 \\

        \textbf{Qwen-3.6} (Local)
        & 0.993 & 0.540 & 0.678
        & 0.920 & 0.932 & 0.929 & 0.930 \\

        \textbf{Qwen-3.7} (API)
        & \textbf{0.997} & \textbf{0.717} & \textbf{0.817}
        & \textbf{0.958} & \textbf{0.966} & \textbf{0.963} & \textbf{0.964} \\

        \bottomrule
    \end{tabular}}
\end{table}

All systems exhibit very high typed-pair precision, with LOKI remaining above $0.98$ and within two percentage points of the direct LLMs. Their main difference is therefore coverage. Relative to LOKI (GPT-OSS), direct Qwen-3.6 improves pair recall by approximately $11\%$, while Qwen-3.7 improves it by approximately $48\%$. This higher coverage propagates to the final Typed Tables, where the direct Qwen-3.6 and Qwen-3.7 baselines achieve approximately $25\%$ and $30\%$ higher macro-F1 than LOKI (GPT-OSS), respectively. In contrast, the two LOKI configurations remain within approximately $2\%$ of each other in table-level F1, indicating that the discovered relational structure can be materialized consistently using different implementations of the replaceable function $G$.

\begin{table}[h]
    \centering
    \caption{Raw counts underlying the Best-Match Typed-Pair and Typed Table materialization results.}
    \label{tab:exp3_raw_counts}
    \resizebox{\columnwidth}{!}{%
    \begin{tabular}{lccccc|ccc}
        \toprule
        System & Pred. & GT & TP & FP & FN & Micro P & Micro R & Micro F1 \\
        \midrule

        \multicolumn{9}{l}{\textit{\textbf{Best-Match Typed-Pair}}} \\

        \textbf{LOKI} (GPT-OSS)
        & 2882 & 6441 & 2823 & 59 & 3618
        & 0.980 & 0.438 & 0.606 \\

        \textbf{LOKI} (Qwen-3.6)
        & 3051 & 6454 & 2990 & 61 & 3464
        & 0.980 & 0.463 & 0.629 \\

        \textbf{Qwen-3.6} (Local)
        & 3103 & 6468 & 3086 & 17 & 3382
        & 0.995 & 0.477 & 0.645 \\

        \textbf{Qwen-3.7} (API)
        & 4149 & 6468 & 4135 & 14 & 2333
        & \textbf{0.997} & \textbf{0.639} & \textbf{0.779} \\

        \midrule

        \multicolumn{9}{l}{\textit{\textbf{Typed Table Materialization}}} \\

        \textbf{LOKI} (GPT-OSS)
        & 2018 & 2018 & 1696 & 322 & 322
        & 0.840 & 0.840 & 0.840 \\

        \textbf{LOKI} (Qwen-3.6)
        & 2011 & 2113 & 1705 & 306 & 408
        & 0.848 & 0.807 & 0.827 \\

        \textbf{Qwen-3.6} (Local)
        & 523 & 523 & 481 & 42 & 42
        & 0.920 & 0.920 & 0.920 \\

        \textbf{Qwen-3.7} (API)
        & 548 & 548 & 525 & 23 & 23
        & \textbf{0.958} & \textbf{0.958} & \textbf{0.958} \\

        \bottomrule
    \end{tabular}}
    \begin{minipage}{0.98\columnwidth}
    \scriptsize
    The raw counts are pooled over each system's evaluable outputs, so GT denominators may differ slightly despite the same 382-admission test set.
    \end{minipage}
\end{table}

LOKI's pair-level errors are dominated by false negatives rather than false positives (Table~\ref{tab:exp3_raw_counts}): 2,823 of its 2,882 predicted pairs are correct, but 3,618 ground-truth pairs remain unrecovered. Qwen-3.7 improves primarily by expanding this coverage, reducing missed pairs by roughly one third rather than by improving an already high precision. At the typed-pair level, LOKI’s main limitation is therefore coverage rather than precision.

\medskip

\stitle{RQ2: How practical is discovery-driven integration?}\\
Table~\ref{tab:exp3_cost} reports the computational and monetary cost of end-to-end integration. LOKI reduces the downstream LLM workload by approximately $67$--$69\%$, requiring only about one third as many tokens as direct prompting. With GPT-OSS as $G$, the complete pipeline is approximately $2\times$ faster than direct local Qwen-3.6 and $3.9\times$ faster than the frontier API baseline. Under the corresponding API-equivalent pricing assumptions, it is also approximately $11\times$ less expensive than direct Qwen-3.6 and $44\times$ less expensive than the frontier API baseline. Comparing LOKI ($G{=}$Qwen-3.6) against direct Qwen-3.6 separates token efficiency from backend latency. LOKI requires approximately $3\times$ fewer LLM tokens and reduces the corresponding token-priced cost by approximately $3.1\times$ relative to direct Qwen-3.6, despite this particular implementation of $G$ taking approximately $4.4\times$ longer. The additional latency therefore stems from the selected downstream implementation rather than a larger LLM reasoning workload.

\begin{table}[h]
    \centering
    \caption{Runtime, LLM token consumption, and estimated API-equivalent cost for Experiment~3.}
    \label{tab:exp3_cost}
    \resizebox{\columnwidth}{!}{%
    \begin{tabular}{lcccc}
        \toprule
        System & Time/Adm. (s) & Tokens/Adm. (K) & Total Tokens (M) & Est. Cost (\$) \\
        \midrule

        \textbf{LOKI} (GPT-OSS)
        & \textbf{44.8} & \textbf{7.2}$^{\dagger}$
        & \textbf{2.75}$^{\dagger}$ & \textbf{$\sim$0.70}$^{\ddagger}$ \\

        \textbf{LOKI} (Qwen-3.6)
        & 391.3 & \textbf{7.2}$^{\dagger}$
        & \textbf{2.75}$^{\dagger}$ & $\sim$2.53$^{\ddagger}$ \\

        \textbf{Qwen-3.6} (Local)
        & 89.9 & 22.1 & 8.43 & $\sim$7.76$^{\ddagger}$ \\

        \textbf{Qwen-3.7} (API)
        & 175.9 & 23.1 & 8.82 & $\sim$30.60$^{\ddagger}$ \\

        \bottomrule
    \end{tabular}}
    \begin{minipage}{0.98\columnwidth}
    \scriptsize
    $^{\dagger}$LOKI tokens cover only the downstream Generalization Function $G$;
    $^{\ddagger}$Costs are estimated over 382 admissions using provider token prices.
    \end{minipage}
\end{table}

\medskip

\stitle{Summary.}
Exp.~3 shows that LOKI transforms discovered row-sentence associations into text-mediated join paths, disambiguates their relationships, and materializes them as provenance-preserving Typed Tables. Direct LLMs achieve stronger final materialization, primarily through higher coverage, whereas LOKI additionally performs the preceding discovery while requiring only about one third of the LLM token workload. By restricting LLM reasoning to compact evidence-backed relationship descriptions, the replaceable function $G$ can be selected according to latency, cost, and deployment requirements.

\section{Conclusion}

This paper studies the discovery of latent relational structure between disjoint tables and unstructured text, uncovering not only which sources are relevant but how they are connected. We formalized \textbf{Text-Mediated Join Path Discovery} as the task of identifying and organizing row-sentence associations into coherent relational structures, and introduced \textbf{LOKI}, a horizontal bidirectional cross-attention model that, from coarse-grained supervision alone, induces fine-grained row-sentence associations, composes them into join paths, and organizes these into relation-consistent groups supporting provenance-aware integration.

Candidate discovery currently requires scoring all table-document pairs, with a cost proportional to $|\mathcal{T}||\mathcal{D}|$; although these evaluations can be batched, an indexable task-specific retrieval representation remains an important direction for scaling to substantially larger data lakes. A second limitation is coverage: LOKI recovers typed row pairs with high precision but lower recall compared to direct LLM prompting, so improving recall of the underlying row-sentence associations is the clearest path to stronger end-to-end integration. A third limitation concerns relationship labeling: as a joint table-text representation model, LOKI delegates canonical relationship naming to the \textit{Generalization Function} $G$. This concerns only downstream ontology-specific labeling, not the discovery of the underlying relational structure; future work aims to improve $G$ via ontology-aware calibration and learnable naming.

Our evaluation across four complementary datasets demonstrates this progression from representation learning to multi-modal discovery and, ultimately, integration, at roughly one third of the LLM token cost of competing approaches. More broadly, unstructured text can serve as explicit relational evidence that makes disconnected data interpretable, auditable, and queryable.




\bibliographystyle{ACM-Reference-Format}
\bibliography{references}

\end{document}